\documentclass[final,3p,times,twocolumn]{elsarticle}

\usepackage[fleqn]{amsmath}
\usepackage{amssymb}

\usepackage{lineno}

\usepackage{placeins}
\usepackage{float}
\usepackage{siunitx}
\usepackage{booktabs}  
\usepackage{multirow}
\usepackage[official]{eurosym}
\usepackage{url}
\usepackage{color}
\usepackage{eurosym}
\usepackage{subcaption}
\usepackage{framed}
\usepackage{nomencl}
\usepackage{multicol}

\makenomenclature

\renewcommand*\nompreamble{\begin{multicols}{2}}
\renewcommand*\nompostamble{\end{multicols}}

\biboptions{longnamesfirst,comma,sort&compress}

\journal{Renewable Energy}

\DeclareSIUnit\uwh{Wh}
\DeclareSIUnit\ukwh{kWh}
\DeclareSIUnit\uco{CO_2}
\DeclareSIUnit\uemi{\kilo\gram\uco\per\ukwh}

\begin{document}

\begin{frontmatter}



\title{Profitability of Photovoltaic and Battery Systems on Municipal Buildings\tnoteref{copyr}}


\author{Rafael Hirschburger\fnref{unr}}
\author{Anke Weidlich\corref{cor1}\fnref{inatech}}
\fntext[unr]{EnBW Energie Baden-W\"{u}rttemberg AG, Germany}
\fntext[inatech]{Department of Sustainable Systems Engineering, University of Freiburg, Germany}
\cortext[cor1]{Corresponding author}
\tnotetext[copyr]{\copyright~2020. This manuscript version is made available under the CC-BY-NC-ND 4.0 license \url{http://creativecommons.org/licenses/by-nc-nd/4.0/}. The text provided here is the same as the accepted manuscript, except for a small correction in the caption of Fig.\,\ref{fig:SSR}, and the addition of tables that provide the numbers for some of the figures (see Appendix). The journal manuscript is available under \url{https://doi.org/10.1016/j.renene.2020.02.077}.}

\date{}

\begin{abstract}
The increasing gap between electricity prices and feed-in tariffs for photovoltaic (PV) electricity in many countries, along with the recent strong cost degression of batteries, led to a rise in installed combined PV and battery systems worldwide. The load profile of a property greatly affects the self-consumption rate and, thus, the profitability of the system. Therefore, insights from analyses of residential applications, which are well studied, cannot simply be transferred to other types of properties. In comparison to residential applications, PV is especially suitable for municipal buildings, due to their better match of demand and supply. In order to analyze the value of additional batteries, municipal PV battery systems of different sizes were simulated, taking load profiles of 101 properties as inputs. It was found that self-consumption differs significantly from households, while different types of municipal buildings are largely similar in terms of the indicators analyzed. The share of electricity consumed during summertime was found to have the most significant impact on the self-consumption rate for most considered system sizes. Due to lower electricity tariffs and lower increases in self-consumption provided through batteries in municipal buildings, the investment into a battery is not economically advantageous in most of the cases considered. 
\end{abstract}


\end{frontmatter}


\begin{table*}[h]   
	\begin{framed}
		\printnomenclature
	\end{framed}
\end{table*}

\section{Introduction}
\label{sec:intro}

In countries in which grid parity has been reached for decentralized photovoltaic (PV) power plants, self-consumption is the main driver for the profitability of such systems. Stationary battery storage can increase the amount of self-consumed electricity significantly. Due to technological improvement and strong cost reductions of lithium-ion batteries, more and more PV systems are nowadays combined with batteries \cite{Figgener2018}.

Current research focuses strongly on storage applications for residential buildings. Results from these studies demonstrate the impact of the load profile on the self-consumption rate (SCR), self-sufficiency rate (SSR) and the profitability of the PV battery system. Therefore, findings cannot directly be applied to other types of buildings and uses. With PV systems nowadays being installed on many schools, daycare centers, town halls and other community buildings, municipal properties are important applications of PV systems. Due to the good temporal match of demand and PV generation in these buildings, self-consumption rates are typically higher than for residential buildings. There are no studies, however, that analyze SCR, SSR and profitability of PV and battery storage systems on municipal properties specifically, and for different battery management strategies. The aim of this study is to close this gap and to provide realistic findings that can be used by project planners, municipal authorities and policy makers. For this purpose, the following research questions are formulated: What SCR and SSR can be realized for different kinds of municipal buildings and sizes of PV storage systems, and which variables influence SCR and SSR the most? Do the current framework conditions allow for an economic operation of PV storage systems on municipal properties? Which are the strongest influencing factors on the profitability? 

Many studies have investigated self-consumption, self-sufficiency and profitability of PV battery storage systems, along with different charging and discharging strategies. The authors of \cite{Tjaden2014} calculate SCR and SSR for different sizes of PV rooftop systems and battery storage, based on a reference household load profile. A model to estimate the internal rate of return for different PV plant and battery sizes, using simulated household load profiles, is developed by \cite{BERTSCH2017}. The authors of \cite{LINSSEN2017} determine the cost optimum constellation of PV system size and storage capacity, by varying both parameters. The authors of \cite{MEREI2016} calculate SSR, SCR and electricity cost for different PV system and battery sizes, based on the load profiles of two supermarkets. The authors of \cite{OLASZI2017} derive equations to calculate SSR and SCR as a function of normalized PV system and storage size by applying an artificial neural network. They also investigate the impact of different discharging strategies on the grid and economic performance of the system. Their results are based on one residential load profile. The authors of \cite{KLINGLER2017} compare a forecast-based operation, based on linear optimization, to self-consumption maximizing operation for residential applications. According to \cite{LINSSEN2017}, there are significant differences in SSR and SCR between simulated, aggregated and standard load profiles. The authors of \cite{Tjaden2014} find that SSR varies by 23\,\% for different households with equal (relative) PV system and storage size. The authors of \cite{NYHOLM2016} determine SCR and SSR for more than 2\,000 Swedish households and find differences in SCR of more than 20\,\% for the same (relative) PV system and storage size. The authors of \cite{Fina2019} focus on the profitability of PV in a municipal setting, but look at the aspect of communities of (mostly residential) buildings rather than municipal buildings. To the best knowledge of the authors, no studies specifically focusing on municipal buildings are yet available.

The conditions of prices, support schemes and solar irradiation for Germany are taken as the basis for the analysis in this study, but general findings can also be transferred to other locations with similar framework conditions. Linear programming is used to simulate two types of battery management systems (BMS), one that minimizes electricity procurement cost and one that additionally keeps the maximum power injection to the public grid as low as possible. The latter aspect is useful for PV support schemes that imply limits on the possible feed-in power. The model inputs are PV generation data for one year with a resolution of 15\,min, system parameters of the PV plant, the battery and the inverter, and 15\,min measured load profiles of 101 municipal properties. The model is run for 110 different PV storage system sizes for each property. SCR, SSR and losses that occur due to different feed-in power limits are calculated. Taking into account electricity prices, feed-in tariffs as well as acquisition and maintenance cost, the internal rate of return (IRR) and the break-even price for the battery at which the IRR would just be zero are calculated. The results are compared to each other, and explanatory variables for potential variances are identified using the generalized linear regression model (GLM) and analysis of variance (ANOVA). 

The remainder of the paper is structured as follows: The applied model, assumptions made and the data input are presented in Sec.~\ref{sec:method}. In Sec.~\ref{sec:results}, results of both models, their statistical evaluation and a sensitivity analysis are presented. Sec.~\ref{sec:discussion} discusses the main findings. Finally, Sec.~\ref{sec:conclusions} concludes the findings.

\section{Method and Data}\label{sec:method}

The programming problem used for minimum cost battery management is presented in Sec.~\ref{sec:model}, and the parameter assumptions and data inputs are described in Sec.~\ref{sec:data}.

\subsection{The Model}\label{sec:model}

Linear programming is used to simulate battery management with two different objectives. The first objective function, $z_1$, minimizes electricity procurement costs, with revenues considered as negative costs. This is represented by the energy procured from the supplier, $E^\text{supply}_t\geq 0~\forall t$, multiplied with the price $p^\text{supply}$, reduced by the feed-in tariff ($FIT$) earned for the energy fed into the public grid, $E^\text{feed-in}_t\geq 0~\forall t$, all summed over the time instances $t=1,...,T$ (Eq.~\ref{eq:cost}). The time horizon is one year, and time increment $\Delta t$ is 15\,min. Under the assumption of a time-invariant tariff and $p^\text{supply}>FIT$, Eq.~\ref{eq:cost} is equivalent to maximizing self-consumption.

\nomenclature{$t \in T$}{Time interval (-)}
\nomenclature{$E^\text{supply}$}{Electrical energy procured from the supplier (MWh)}
\nomenclature{$E^\text{feed-in}$}{Electrical energy fed into the public grid (MWh)}
\nomenclature{$FIT$}{Feed-in tariff (EUR/MWh)}
\nomenclature{$z_1$}{Objective function 1 (cost minimization)}
\nomenclature{$p^\text{supply}$}{Electrical energy procurement price (EUR/MWh)}
\nomenclature{$\Delta t$}{Time increment (min)}

\begin{equation} \label{eq:cost}
\min z_1 = \sum_{t=1}^T \left(E^\text{supply}_t\cdot p^\text{supply} - E^\text{feed-in}_t\cdot FIT\right)
\end{equation}

This operation strategy is observed to provide only limited grid relief (in terms of highest power injected into the public grid), because the battery is often filled before the daily peak of PV generation is reached \cite{MOSHOVEL2015,RANAWEERA2016}. As a result, several PV support schemes have implemented some limit on power injection to the grid, as this has proven to be an effective measure for limiting grid interaction \cite{Sato2019}. If the maximum annual power injection is entered into the objective function, battery capacity can be saved until the time of maximum PV production, and energy above the feed-in power limit can be stored \cite{RANAWEERA2016,RUF2018}. The second objective function, $z_2$, therefore minimizes both cost and peak power feed-in, $P^\text{feed-in,max}\geq 0$ (Eq.~\ref{eq:costpower}). Both parts of the objective functions are normalized and weighted. The normalizing term $\Omega$ represents the cost of buying the complete demand from the grid, and $P^\text{PV}$ represents the installed PV generation capacity. Normalization turns the cost and power part into dimensionless expressions.

\nomenclature{$P^\text{feed-in,max}$}{Peak power feed-in (MW)}
\nomenclature{$z_2$}{Objective function 2 (minimization of cost and power feed-in)}
\nomenclature{$\lambda$}{Objective weighting factor (-)}
\nomenclature{$\Omega$}{Cost normalization factor (EUR)}

\begin{equation} \label{eq:costpower}
\begin{split}
\min z_2 & = \lambda \cdot \sum_{t=1}^T  \frac{E^\text{supply}_t\cdot p^\text{supply} - E^\text{feed-in}_t\cdot FIT}{\Omega} \\
& = +(1-\lambda)\cdot \frac{P^\text{feed-in,max}}{P^\text{PV}}
\end{split}
\end{equation}

Weighting is done through the weighting factor $\lambda$, with $0\leq \lambda\leq 1$. As neither the electricity tariff nor the FIT is assumed to be time-varying, some degree of grid feed-in reduction can be achieved through shifting the timing of battery charging without deviating from the cost minimum from $z_1$. The $\lambda$ value is set so that this cost minimum is achieved.

The building's electricity consumption in one time interval, $E^\text{demand}_t$, and its PV production, $E^\text{PV}_t$ both go into the model for the whole year, so perfect foresight is assumed. Values are taken from historic measured data of 2017. The energy balance constraint is given by Eq.~\ref{eq:balance}, with electricity charged to and discharged from the battery represented by $E^\text{ch}_t\geq 0~\forall t$ and  $E^\text{dch}_t\geq 0~\forall t$, respectively.

\nomenclature{$E^\text{demand}$}{Electricity consumption (MWh)}
\nomenclature{$E^\text{PV}$}{Electricity generation from a photovoltaic plant (MWh)}
\nomenclature{$E^\text{ch}$}{Electricity charged to the battery (MWh)}
\nomenclature{$E^\text{dch}$}{Electricity discharged from the battery (MWh)}
 
 \begin{equation} \label{eq:balance}
E^\text{demand}_t - E^\text{PV}_t= E^\text{supply}_t - E^\text{feed-in}_t + E^\text{dch}_t - E^\text{ch}_t ~~\forall t
\end{equation}

The battery state of charge $SOC_t$ at time $t$ can be in a range up to its energy capacity $C$ (\ref{eq:SOC1}), and is defined by (\ref{eq:SOC2}). It considers charging and discharging efficiencies $\eta^\text{ch}$ and $\eta^\text{dch}$, respectively, with $0\leq\eta^\text{ch},\eta^\text{dch}\leq 1$.

 \nomenclature{$SOC$}{Battery state of charge (MWh)}
 \nomenclature{$C$}{Battery storage capacity (MWh)}
 \nomenclature{$\eta^\text{ch}$}{Battery charging efficiency (-)}
 \nomenclature{$\eta^\text{dch}$}{Battery discharging efficiency (-)}

  \begin{equation} \label{eq:SOC1}
 0\leq SOC_t\leq C~~\forall t
 \end{equation}
 
  \begin{equation} \label{eq:SOC2}
SOC_t= SOC_{t-1}+\eta^\text{ch}\cdot E^\text{ch}_t-\frac{1}{\eta^\text{dch}}\cdot E^\text{dch}_t~~\forall t
 \end{equation}
 
  Charging and discharging power is limited by maximum power values $P^{\text{ch,max}}$ and $P^{\text{dch,max}}$, respectively. These maximum power values are expressed in relation to the charging capacity $C$ of the battery, through the maximum power rates $r^\text{ch,max}$ and $r^\text{dch,max}$ (in MW/MWh).

  \nomenclature{$P^{\text{ch,max}}$}{Maximum charging power (MW)}
  \nomenclature{$P^{\text{dch,max}}$}{Maximum discharging power (MW)}
  \nomenclature{$r^{\text{ch,max}}$}{Maximum charging power rate (MW/MWh)}
  \nomenclature{$r^{\text{dch,max}}$}{Maximum discharging power rate (MW/MWh)}
 
 \begin{equation}
\frac{E^{\text{ch}}_t}{\Delta t}\leq P^{\text{ch,max}} = r^\text{ch,max}\cdot C~~\forall t
 \end{equation} 
 
 \begin{equation}
\frac{E^{\text{dch}}_t}{\Delta t}\leq P^{\text{dch,max}} = r^\text{dch,max}\cdot C~~\forall t
 \end{equation} 
  
 The ambition to keep the maximum annual power feed-in into the grid as low as possible is achieved by the constraint defined through (\ref{eq:peakpower}). In this, the peak feed-in power is calculated as the average power over the 15\,min time interval with highest feed-in energy of the year.
 
   \begin{equation} \label{eq:peakpower}
\frac{E^\text{feed-in}_t}{\Delta t} \leq  P^\text{feed-in,max} ~~\forall t
 \end{equation}
 
 In order to only charge in situations of local surplus, and only discharge in situations of local net demand, the constraints given by (\ref{eq:charge}) and (\ref{eq:discharge}) apply. These are only necessary if $z_2$ applies. Although it would be expected that (dis)charging (to) from the grid should be discouraged by the fact that (dis)charge efficiencies are taken into account in this model, it has been observed that this is not the case for all $\lambda$ values. If a high weight is applied to minimizing the PV feed-in, frequent charging and discharging happens as a strategy to discard energy that would otherwise be injected to the grid. However, if (\ref{eq:charge}) and (\ref{eq:discharge}) are introduced, cost minimal outputs with low PV feed-in and without "wasting" energy through frequent charging and discharging can be found for a wide range of $\lambda$ values in all considered cases.
 
   \begin{equation} \label{eq:charge}
   E^\text{ch}_t (E^\text{demand}_t - E^\text{PV}_t)\leq 0 ~~\forall t
\end{equation}
 
   \begin{equation} \label{eq:discharge}
   E^\text{dch}_t (E^\text{PV}_t - E^\text{demand}_t)\leq 0 ~~\forall t
   \end{equation}
 
 Using the simulated variables, self-consumption rate SCR and self-sufficiency rate SSR are calculated using Eqs.~\ref{eq:SCR} and \ref{eq:SSR}. The calculation also reflects losses from charging and discharging the battery.
 
   \begin{equation} \label{eq:SCR}
 SCR = \frac{\sum_t \left(E^\text{PV}_t-E^\text{feed-in}_t-\left(1-\eta^\text{ch}\right)E^\text{ch}_t-\left(1-\frac{1}{\eta^\text{dch}}\right)E^\text{dch}_t\right)}{\sum_t E^\text{PV}_t}
 \end{equation}
 
   \nomenclature{$SCR$}{Self-consumption rate (-)}
   \nomenclature{$SSR$}{Self-sufficiency rate (-)} 
 
    \begin{equation} \label{eq:SSR}
 SSR = \frac{\sum_t \left(E^\text{PV}_t-E^\text{feed-in}_t-\left(1-\eta^\text{ch}\right)E^\text{ch}_t-\left(1-\frac{1}{\eta^\text{dch}}\right)E^\text{dch}_t\right)}{\sum_t E^\text{demand}_t}
 \end{equation}
 
 For the profitability calculations, the sum of all power feed-ins above the threshold of 70\,\% of the respective nominal PV capacity is calculated for each PV battery system. This is inspired by current regulation of the feed-in tariff in Germany, which limits grid injections to 70\,\% of the PV capacity. The energy losses related to this regulation are integrated into the profitability evaluation. 
 
 Profitability of a PV battery system is best assessed through a discounted cash flow method \cite{weidlich2012decentralized}. The specific indicator chosen here is the (real) internal rate of return (IRR) as defined by Eq.~\ref{eq:IRR}, in which $I_0$ quantifies the initial investment, and $c^\text{op}$ the yearly operating costs. It has been calculated using the respective Matlab function.
 
     \begin{equation} \label{eq:IRR}
 0 = -I_0 + \sum_{t=1}^T \frac{c^\text{op}}{(1+IRR)^t}
 \end{equation}
 
    \nomenclature{$I_0$}{Initial investment expenditure (EUR)} 
    \nomenclature{$c^\text{op}$}{Operating expenditures (EUR/a)} 

 Both models (using $z_1$ and $z_2$, respectively) are run for all properties for which historical data is available, and for 110 hypothetical system configurations regarding PV size and battery size. In order to make results comparable, system sizes are always given in relation to the yearly electricity consumption of each property, in kW$_\text{p}$/MWh for the PV generation capacity, and kWh/MWh for battery storage capacity. Ten different PV sizes from 0.2 to 2\,kW$_\text{p}$/MWh and eleven battery sizes from 0 to 2\,kWh/MWh are analyzed.


\subsection{Data Input}\label{sec:data}

The technical parameters and cost assumptions used for the battery are based on data listed from an extensive monitoring program of actual system implementations \cite{Figgener2018} and from market observations \cite{CARMEN}; the assumed battery investment includes inverter costs. Prices for PV systems are based on \cite{Wirth2019}. Feed-in tariffs for different PV sizes, $P^\text{PV}$ (installed power, in kW), are assumed as granted in Germany in 2019.\footnote{In Germany, only PV plants up to a size of 100\,kW$_\text{p}$ get the feed-in tariff; larger plants can receive a premium on top of their market revenues. For simplicity, it is assumed here that the sum of market revenue and premium amounts to the same level as the $FIT_3$ given in Tab.~\ref{tab:assumptions}.} Electricity prices for municipalities are based on a survey of actual contracts that were available to the authors in the course of this study. In line with current German regulation, a surcharge rate of 27.5\,EUR/MWh is due if electricity is used from an own PV plant, if that plant is supported through a feed-in tariff. An overview of all assumptions made is given in Tab.~\ref{tab:assumptions}.
 
    \nomenclature{$P^\text{PV}$}{Photovoltaic installed capacity (kW)} 
 
\begin{table*}[h]
	\centering
	\begin{tabular}{llrrl}
		\hline
		\multicolumn{2}{l}{\textbf{Parameter}} & & \textbf{Value} & \textbf{Unit} \\
		\hline
	\multicolumn{2}{l}{Charging / discharging efficiency} & $\eta^\text{ch},\eta^\text{dch}$	& 0.94 & $-$ \\
	\multicolumn{2}{l}{Maximum charging power rate} & $r^\text{ch,max}$ & 0.6 & MW/MWh \\
	\multicolumn{2}{l}{Maximum discharging power rate} & $r^\text{dch,max}$ & 0.6 & MW/MWh \\
	\multicolumn{2}{l}{Electricity tariff} & $p^\text{supply}$ & 240 & EUR/MWh \\
	Feed-in tariff & $P^\text{PV} \leq$ 10\,kW$_\text{p}$ & $FIT_1$ & 101.8 & EUR/MWh \\  
    & $10<$ $P^\text{PV} \leq$ 40\,kW$_\text{p}$ & $FIT_2$ & 99.0 & EUR/MWh \\  
	& $P^\text{PV} > 40$ & $FIT_3$ & 77.8 & EUR/MWh \\  
	Market revenue &  & MR & 40.0 & EUR/MWh \\
	\multicolumn{2}{l}{PV investment} & $I_0^\text{PV}$ & 1\,150 & EUR/kW$_\text{p}$ \\
	\multicolumn{2}{l}{Battery investment} & $I_0^\text{batt}$ & 800 & EUR/kWh \\
	\multicolumn{2}{l}{Maintenance cost} & $c^\text{op}$ & 1 & \% of $I_0$/a \\
	\multicolumn{2}{l}{Lifetime} & $T^{\text{PV,batt}}_\text{life}$ & 20 & a \\
	\multicolumn{2}{l}{Weighting factor} & $\lambda$ & 0.01 &$-$\\
	\hline
\end{tabular}
\caption{Parameter values assumed in the case study}
\label{tab:assumptions}
\end{table*}

    \nomenclature{$MR$}{Market revenue (EUR/MWh)} 
    \nomenclature{$T_\text{life}$}{Expected technical lifetime of a plant (a)} 

The load profiles of 101 municipal properties located in the city of Frankfurt and the surrounding area have been downloaded from \cite{FFM}. Each dataset contains the electricity consumption in 15\,min intervals for 2017. 76 of the datasets are from school buildings. On 15 of these properties, daycare centers are located besides the school, whose electricity consumption is measured by the same meter. 45 properties include school buildings and a sports hall. On five of these properties, a second meter is installed, allowing the differentiation between electricity consumed by the sports hall and the school building. Thus, the number of datasets including only a sports hall and only a school could be increased to six and 20, respectively, and the total number of data sets to 111. Another 15 properties contain school buildings as well as sports halls and daycare centers, whose electricity consumption could not be divided. Furthermore, the load profiles of twelve administration buildings, four nursing homes, two museums and seven conference halls have been analyzed. The 76 properties with school buildings can further be divided into seven types of schools: vocational (12), comprehensive (3), elementary (25), high school (11), secondary school (6) and special school (3). In 16 cases, more than one type of school was located on the property.

The yearly electricity consumption of the properties in the datasets ranges from 17\,MWh to 1.2\,GWh. 48.6\,\% consumed less than 100\,MWh, and 73\,\% consumed less than 200\,MWh. Only two buildings (a huge conference hall and an administration building) had an electricity consumption of more than 600\,MWh. Electricity consumption during summer was less than during winter in 100 properties. The share of summertime consumption, which is here defined as the consumption that takes place between April 1$^\text{st}$ and September 30$^\text{th}$, ranges between 19.5 and 57.2\,\%, with an average of 42.5\,\% and quartiles of 39 and 45\,\%.

To simulate electricity generation, the power output data from a PV plant installed on a town hall in a community in the Southwest of Germany (Rastatt) was obtained from the monitoring portal. The nominal capacity of the plant is 37.8\,kW$_\text{p}$, and the production in 2017 was 37\,349\,kWh (988\,kWh/kW$_\text{p}$), which is a typical value for Southern Germany. Generation of preceding years differed by $+3.9$ to $-2.3$\,\%. 60 of the PV modules of the given plant are oriented towards West, and 80 modules face East with and inclination angle of 20$^{\circ}$ and 25$^{\circ}$, respectively. The data is available in 15\,min time resolution. The given PV plant is combined with batteries that follow a conventional control strategy, i.~e. cost minimization without minimization of maximum annual power injection. It could therefore be used for validating the model output using objective function $z_1$ (\ref{eq:cost}), comparing it to the operation observed in real life. It was observed that the model output was very close to the actually observed values, with differences in SCR of 0.83\,\% (34.67\,\% compared to 35.50\,\%).

\section{Results}\label{sec:results}

The linprog function of Matlab was used to calculate all results for each of the 110 system configurations applied to the available $n=111$ consumption data sets. Results are presented for the self-consumption rate in Sec~\ref{sec:SCR}, for the self-sufficiency rate in Sec.~\ref{sec:SSR}, and for profitability in Sec.~\ref{sec:IRR}. 

\subsection{Self-Consumption Rate}\label{sec:SCR}

\begin{figure}[h]
	\centering
	\includegraphics[width=\linewidth]{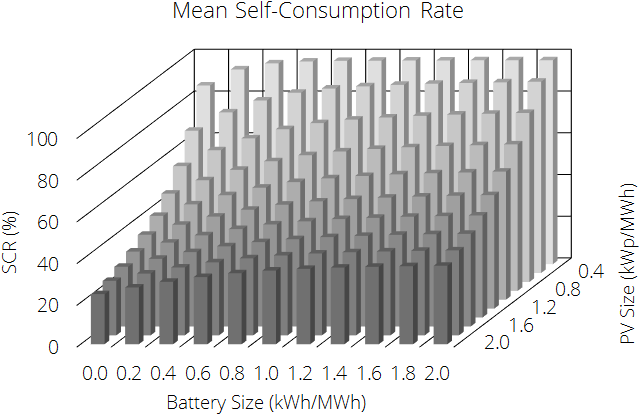}
	\caption{Average SCR of all municipal buildings as a function of PV and battery sizes}
	\label{fig:SCR}
\end{figure}

Fig.~\ref{fig:SCR} shows the average self-consumption rate of all analyzed data sets, for each of the PV and battery sizes assumed. Note that the PV size axis goes from highest to the lowest values. As expected, the SCR decreases with increasing PV size, and increases with rising battery capacity. The standard deviation of SCR values within one configuration is comparably small and nearly proportional to the arithmetic mean of SCR for most system sizes, with a coefficient of variation  between 7.1 and 13.5\,\% for all sizes of PV plant greater than 0.2\,kW/kW$_\text{p}$. 

\begin{figure}[h]
	\centering
	\includegraphics[width=\linewidth]{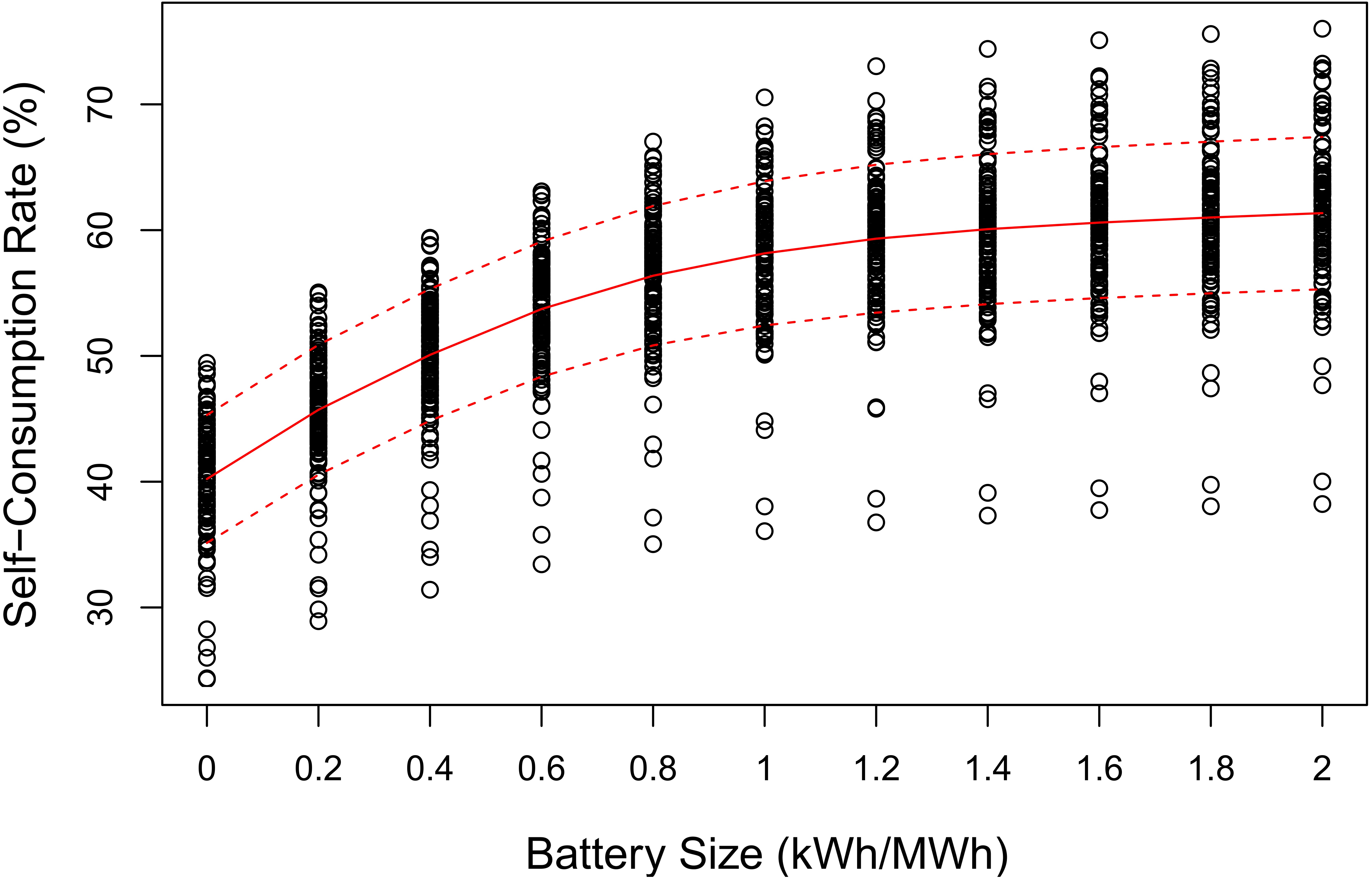}
	\caption{SCR as a function of the battery size, for a 1\,kW$_\text{p}$/MWh PV plant; solid line: average SCR; dashed lines: average SCR plus/minus one standard deviation}
	\label{fig:SCR_Size}
\end{figure}

Fig.~\ref{fig:SCR_Size} shows all values of SCR for a PV system size of 1\,kW$_\text{p}$/MWh, along with the average value and the standard deviation. It can be seen that the increase in SCR flattens for battery capacities greater than 1\,kWh/MWh, and the added value of more kWh of storage capacity becomes comparatively low. 

In order to understand the differences in SCR between the different types of building, electricity consumption, share of summertime and daytime consumption and -- in the case of school buildings -- school types, these variables were considered as features for the variables SCR, SSR and profitability in an analysis of variance (ANOVA) and a generalized linear model (GLM), using the statistics software R. ANOVA is conducted on each single predictor variable to measure the variance explained by the respective variable (R$^2$)  and to test if the predictor variable has a significant impact on the response variable (F-Test). GLM is used to capture the impacts of all relevant variables in one model. GLM allows to identify significant differences between single categories of one predictor variable (t-test). The total variance explained by the model can be calculated from the output. 

The model used to analyze the variance of SCR is given by Eq.~\ref{eq:SCRANOVA}, where $d_{B1}$ is the dummy variable related to building type 1 and similar for the other building types, $EC$ is the electricity consumption, $SC$ is the summertime consumption (as defined previously) and $DC$ is the daytime consumption, which is defined as the fraction of the consumption that takes place between 08:00~a.m. and 08:00~p.m.

\begin{equation}\label{eq:SCRANOVA}
\begin{split}
SCR\sim & N(\mu=\beta_0+\beta_{B1} d_{B1}+\beta_{B2} d_{B2}+\dots +\beta_{EC} EC \\
&+ \beta_{SC} SC + \beta_{DC} DC,\sigma)
\end{split}
\end{equation}

    \nomenclature{$\beta$}{Coefficients of the generalized linear model (-)} 
    \nomenclature{$d_{B}$}{Dummy variable for a building type (-)} 
    \nomenclature{$SC$}{Share of summertíme consumption (-)} 
    \nomenclature{$EC$}{(Yearly) electricity consumption of a building (MWh)} 
    \nomenclature{$DC$}{Share of daytime consumption (-)}

Even though the SCRs of some system sizes are not unambiguously normally distributed, the best fit was reached using a normal distribution (considerably lower AIC\footnote{The Akaike information criterion (AIC) is and index used to compare different statistical models (e.\,g. different distributions or variables).}  compared to log-normal or gamma distribution in all cases). On average, the model was able to explain 78.6\,\% of the initial variance. In general, the explained variance is higher for larger systems (both PV plant and battery). For systems with a PV size of 0.2\,kW$_\text{p}$/MWh, the model could only explain 30.6 to 51.4\,\% of the variance. Correlations between numerical variables (daytime, summertime and total electricity consumption) are low, with Pearson’s $r$ being 0.4 or lower. School buildings tend to have a lower summertime consumption than other types of building and a slightly higher daytime consumption. Yet, the type of school does not seem to be correlated to any other variable.

\begin{figure}[h] 	
	\centering
	\begin{subfigure}[b]{0.64\linewidth}
		\includegraphics[width=\linewidth]{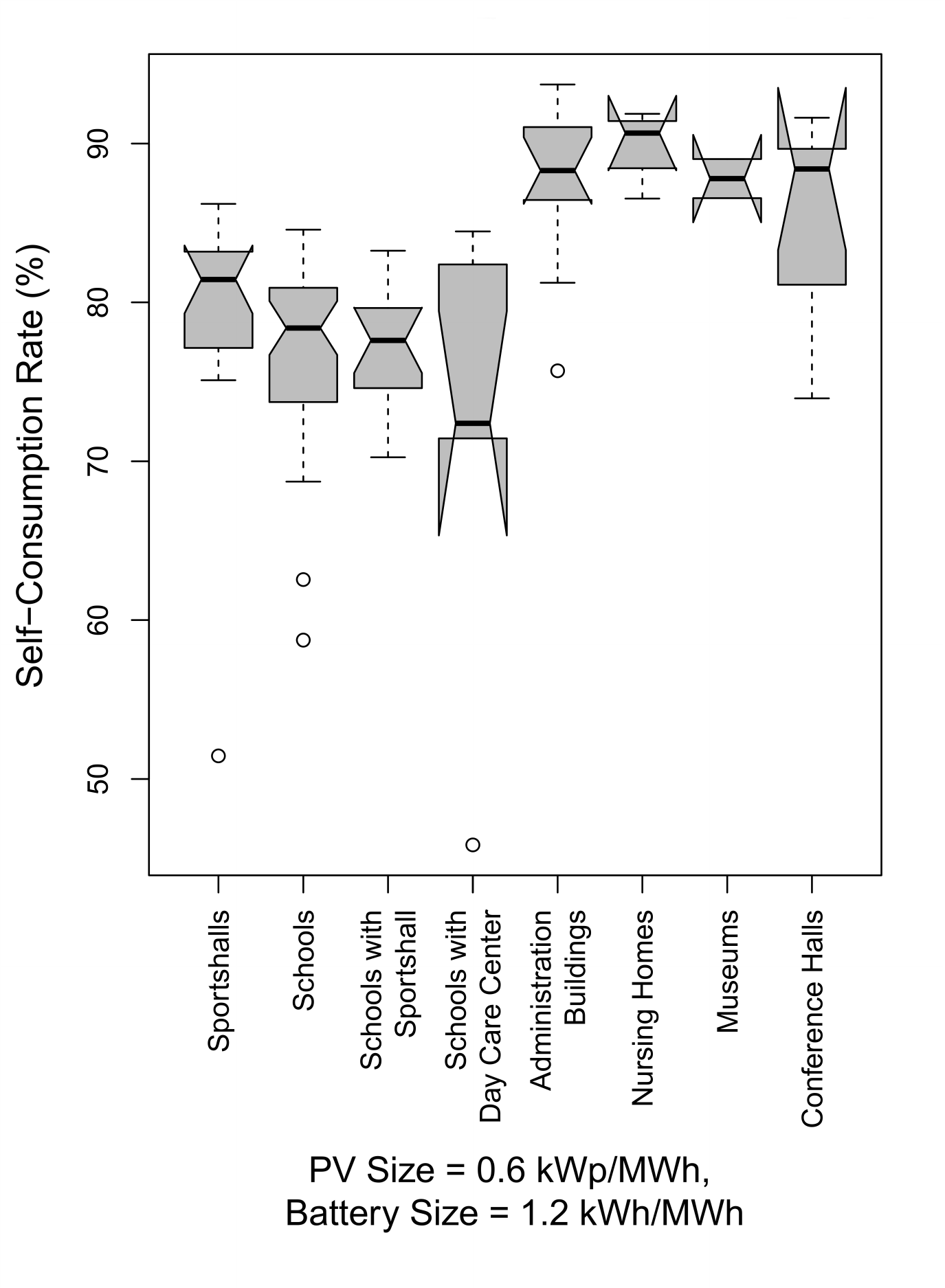} 
		\label{fig:SCR_Summertime}
	\end{subfigure} 
	\begin{subfigure}[b]{0.64\linewidth}
		\includegraphics[width=\linewidth]{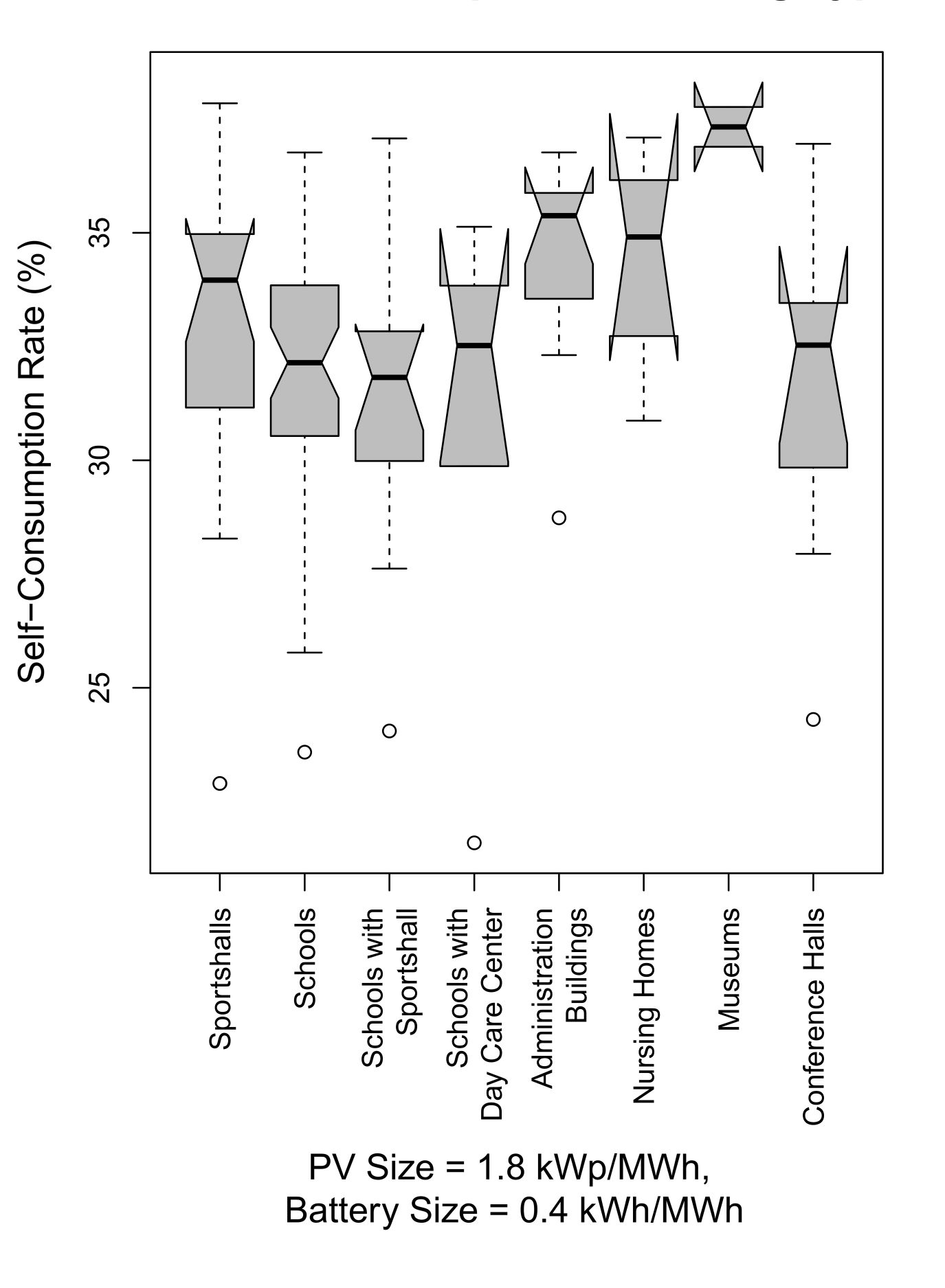}
	\end{subfigure} 
	\caption{Differences in SCR between building types for two different system sizes}
	\label{fig:SCR_Statistics}
\end{figure}

ANOVA shows that the type of building has a significant impact on the self-consumption rate in case of 105 out of 110 system sizes. The variable explains on average 25.7\,\% of the variance. In general, $R^2$ increases with increasing battery size and decreases with PV system size. This can be seen in Fig.~\ref{fig:SCR_Statistics}, which shows overlapping notches of boxplots for most building types for a system with a PV plant of 1.8\,kW$_\text{p}$/MWh and a battery with 0.4\,kWh/MWh, but more noticeable differences for a system with 0.6\,kW$_\text{p}$/MWh and 1.2\,kWh/MWh, respectively. $t$-tests of the linear model show that the SCR of nursing homes, museums and conference halls are significantly different from the SCR of schools in 21, 49 and 57 cases, respectively. The SCR of administration halls, schools with sports hall and schools with daycare center are not significantly different in any case and SCR of sports halls in only four cases. The beta coefficients of the linear model are on average $+2\,\%$ for nursing homes, $+3.2\,\%$ for museums and $-2\,\%$ for conference halls compared to schools without sports hall or daycare center. 


Summertime consumption is found to be the most influential variable effecting SCR. On average, it explained 67.7\,\% of the variance, with up to 90\,\% for systems with large battery and PV system, but lower $R^2$ for small systems. Both $t$-test of GLM and $F$-test of ANOVA confirm a significant impact of summertime consumption for all 110 systems sizes. Fig.~\ref{fig:SCR_Summertime} illustrates the differences in scattering for different system sizes. Beta coefficients show that SCR increases on average by 0.65\,\% for each percent of summertime consumption. The highest impact of 1.07\,\% is found for a system of 0.6\,kW$_\text{p}$/MWh and 2\,kWh/MWh. The lowest impact of 0.25\,\% was observed for a system with 2\,kW$_\text{p}$/MWh and no battery. 

\begin{figure}[h] 	
		\centering
	\begin{subfigure}[b]{0.75\linewidth}
		\includegraphics[width=\linewidth]{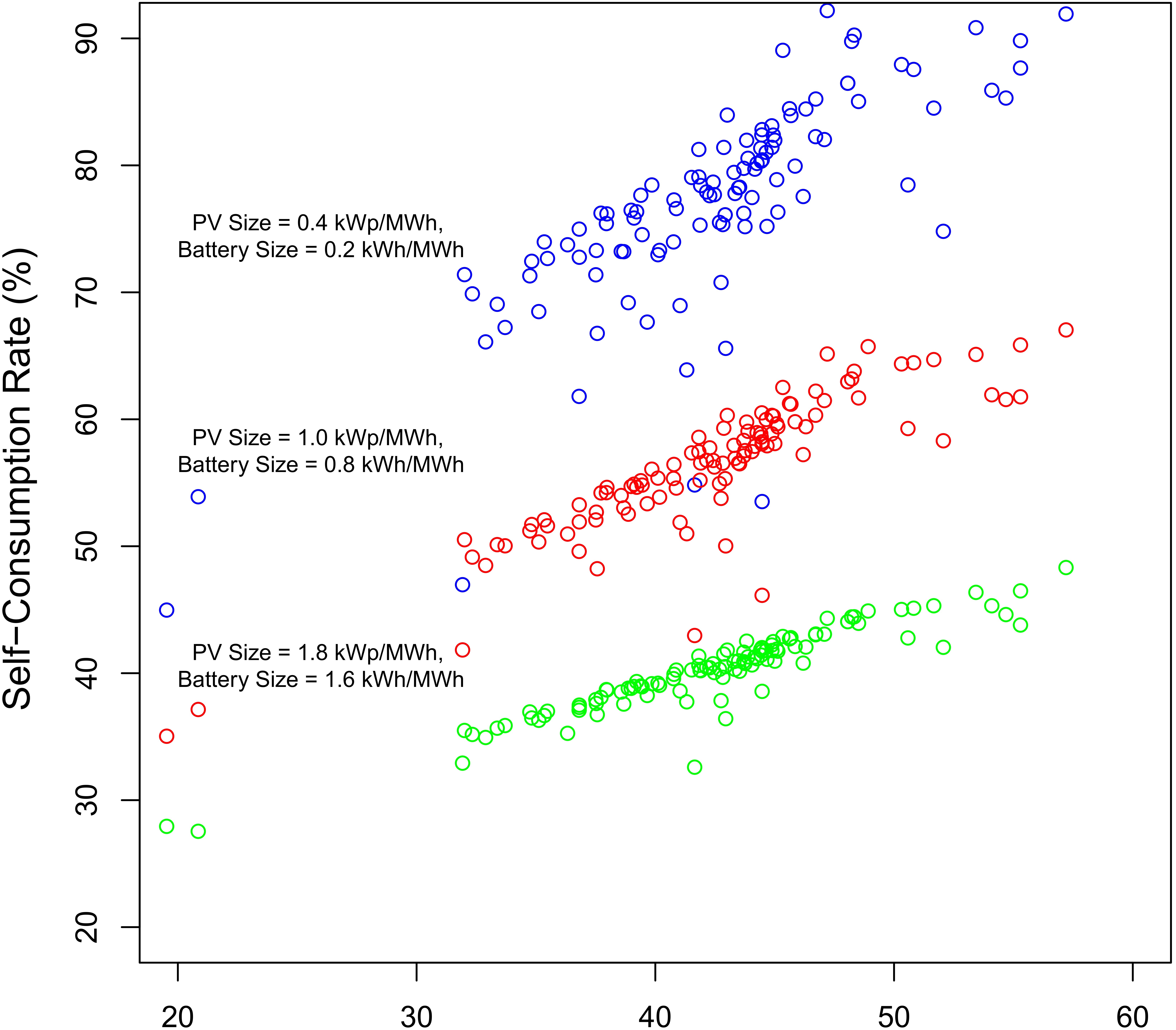} 
		\caption{Summertime consumption (in \%)} 
		\label{fig:SCR_Summertime}
	\end{subfigure} 
	\begin{subfigure}[b]{0.75\linewidth}
		\includegraphics[width=\linewidth]{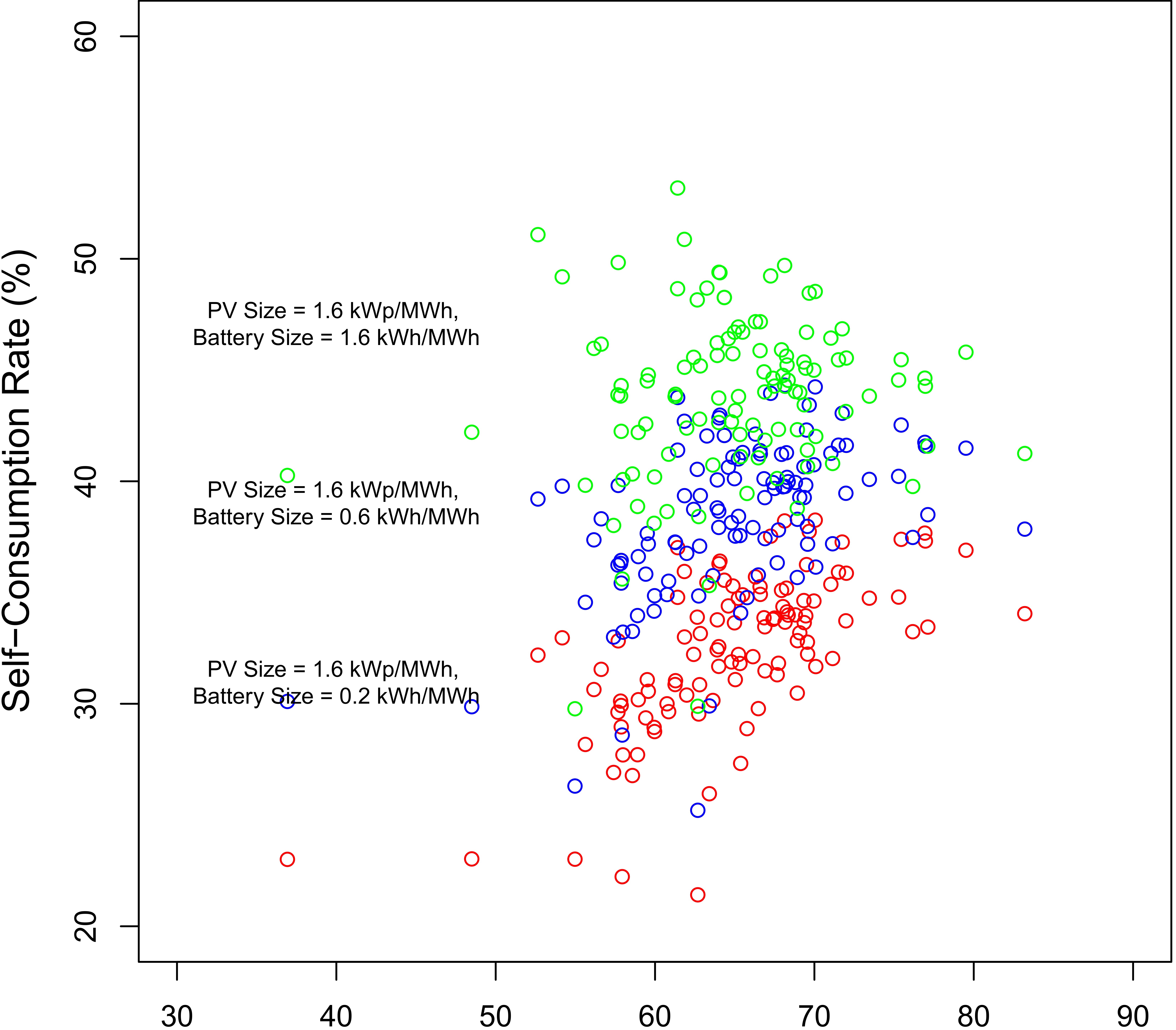}
		\caption{Daytime consumption (in \%)} 
		\label{fig:SCR_Daytime}
	\end{subfigure} 
	\caption{Impact of demand patterns on self-consumption rate}
\end{figure}

Electricity consumption explains on average 14.8\,\% of the initial variance. ANOVA shows a significant impact in 101 cases ($F$-test), and $t$-test of the linear model in 62 cases. Explained variance ranges from 10 to 20\,\% for most system sizes. The highest beta coefficient of GLM was 0.0061\,\% increase of SCR per MWh of yearly electricity consumption. 

The share of daytime consumption is found to have a significant impact on SCR in 50 (ANOVA) and 60 (GLM) cases, and explains on average 10.5\,\% of the total variance. The variance explained by daytime consumption varies extremely between system sizes. For systems with large PV plants and small batteries, the share of daytime consumption explains up to 50\,\%, but for any system with a battery size of more than 0.8\,kWh/MWh, the explained variance is smaller than 13\,\%. Beta coefficients for systems with PV plants greater than 0.6\,kW$_\text{p}$/MWh and batteries smaller than 1\,kWh/MWh are between 0.1 and 0.35\,\% increase in SCR for each percent of daytime consumption. Fig.~\ref{fig:SCR_Daytime} shows that SCR increases with increasing daytime consumption in case of a large PV plant combined with a small battery, but for systems with large PV plants and large batteries, no impact of daytime consumption on SCR can be observed.

\subsection{Self-Sufficiency Rate}\label{sec:SSR}

\begin{figure}[h]
	\centering
	\includegraphics[width=\linewidth]{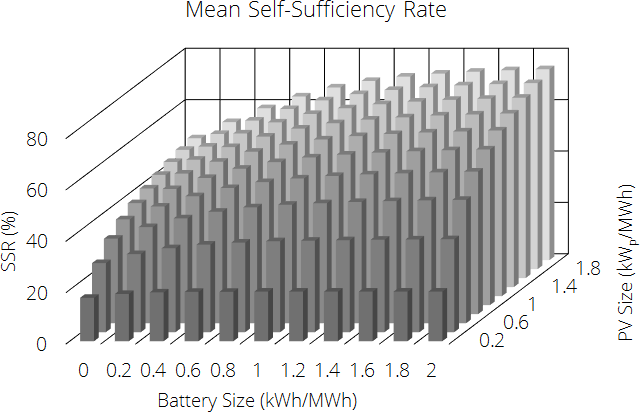}
	\caption{Average SSR of all municipal buildings as a function of PV and battery sizes}
	\label{fig:SSR}
\end{figure}

The average self-sufficiency rate is shown in Fig.~\ref{fig:SSR}. Since SSR is the ratio of self-consumed PV-generated electricity and total electricity consumption, it increases with increasing battery capacity and increasing PV system size. For battery capacities greater than 1\,kWh per kW$_\text{p}$ PV capacity, the increase in SSR is relatively low. Scattering of SSR is similar to the variation of SCR, with coefficients of variation between 7 and 13\,\% for most system sizes. 

Since SCR and SSR are indicators with the same numerator (self-consumed PV electricity), they are perfectly correlated, and properties with a high SCR also have a high SSR. Thus, the variation between properties can be explained by the same model as used for SSR, showing the same $R^2$ for the respective variables. The beta coefficients for the variables are different, however, since the values of SSR are small for small PV plants and large for large PV plants. The beta coefficients for daytime consumption range from 0.1 to 0.5\,\% increase in SSR for each percent of daytime consumption for systems with PV plants greater than 0.6\,kW$_\text{p}$/MWh, and batteries smaller than 1\,kWh/MWh. Beta coefficients for summertime consumption are highest for systems with large PV plants and large battery (about  $+0.9\,\%$ SSR for each percent of summertime consumption) and lowest for systems with small PV plant and small battery ($+0.2$ to $+0.5\,\%$ SSR for each percent of summertime consumption). Beta coefficients for building types are $+1.7$, $+2.9$ and $-2$ for nursing homes, museums and conference halls on average.

\subsection{Profitability}\label{sec:IRR}

\begin{figure}[h]
	\centering
	\includegraphics[width=\linewidth]{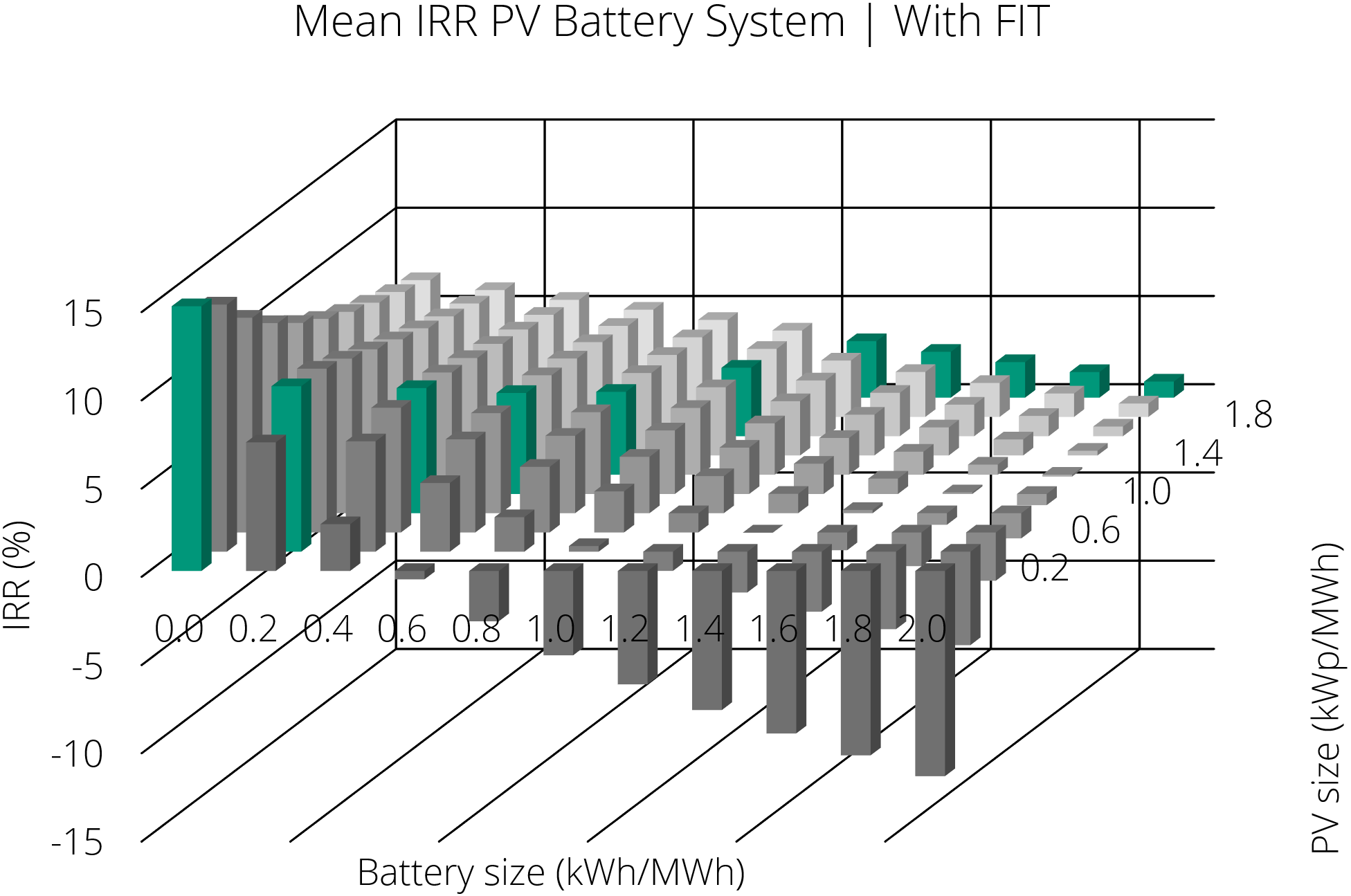}
	\caption{Internal rate of return as a function of PV and battery sizes, assuming FIT}
	\label{fig:IRR_PVBatt_FIT}
\end{figure}

Whether or not a system is seen as profitable for a given internal rate of return depends, of course, on the interest rate aspiration of the investor, and no absolute threshold is proposed here. However, a negative IRR is a clear sign for lacking profitability. IRR values were calculated for all properties and all sizes of PV storage systems. Average values are shown in Fig.~\ref{fig:IRR_PVBatt_FIT} for all PV and battery sizes, assuming the current feed-in tariff scheme in Germany (cp. Tab.~\ref{tab:assumptions}). The green bars indicate the PV size reaching the highest IRR for a given battery size. The highest IRR of 16.3\,\% was found for small PV systems without a battery. It was also found that the IRR decreases with increasing battery capacity for all PV sizes. As small PV systems reach higher SCR values, profitability also decreases with increasing PV size in the FIT scheme. It must be noted that, as the assumed cost of installation is fixed per kW$_\text{p}$ of PV and per kWh of battery, economies of scale for larger PV plants are neglected, here. Another pattern observed in the IRR results is that larger batteries also increase the PV size that makes the system most profitable for the particular battery size. 

\begin{figure}[h]
		\centering
	\begin{subfigure}[b]{\linewidth}
		\includegraphics[width=\textwidth]{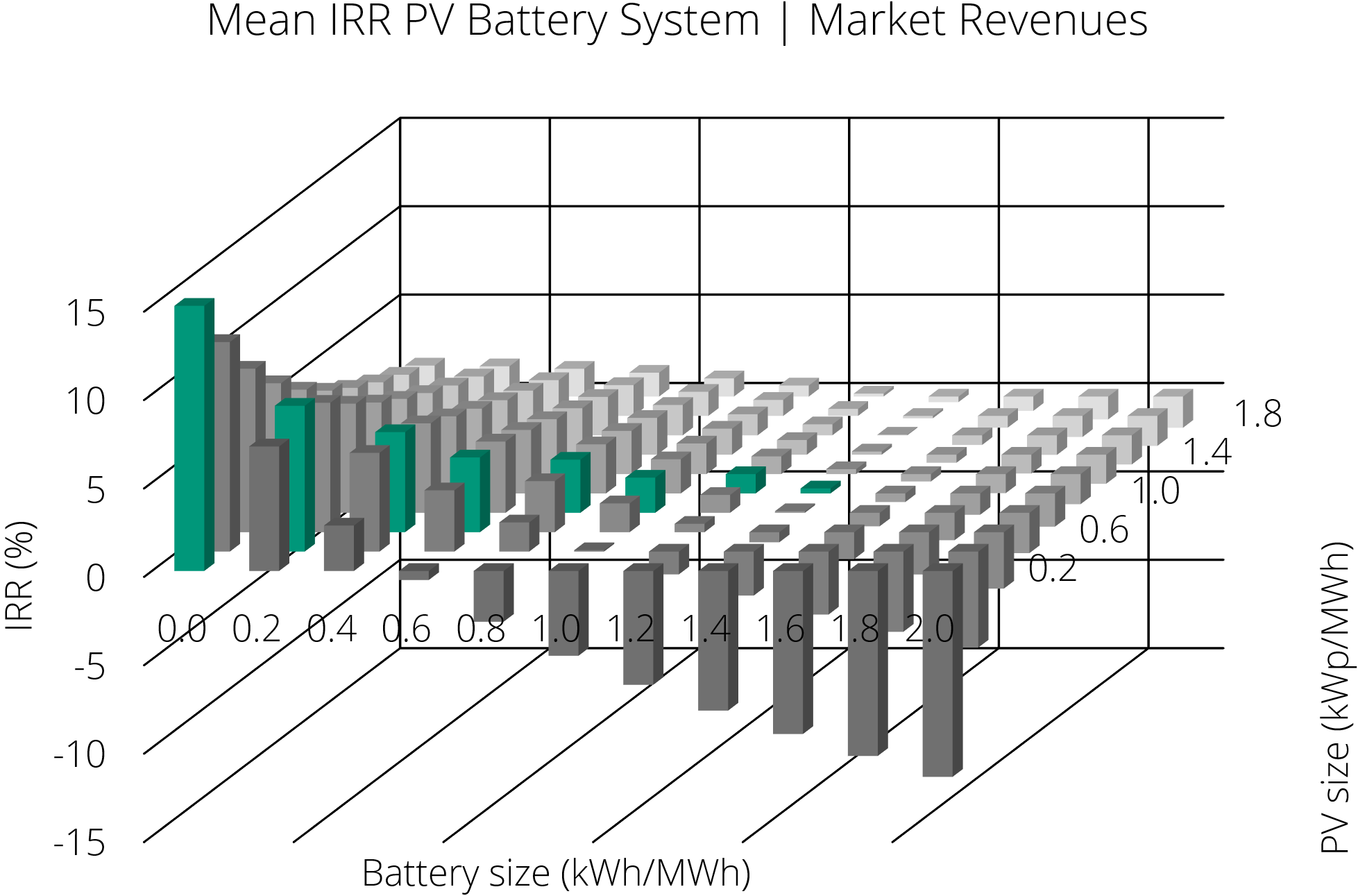}
		\caption{Only market remuneration}
		\label{fig:IRR_PVBatt_Market}
	\end{subfigure} 

	\begin{subfigure}[b]{\linewidth}
		\includegraphics[width=\textwidth]{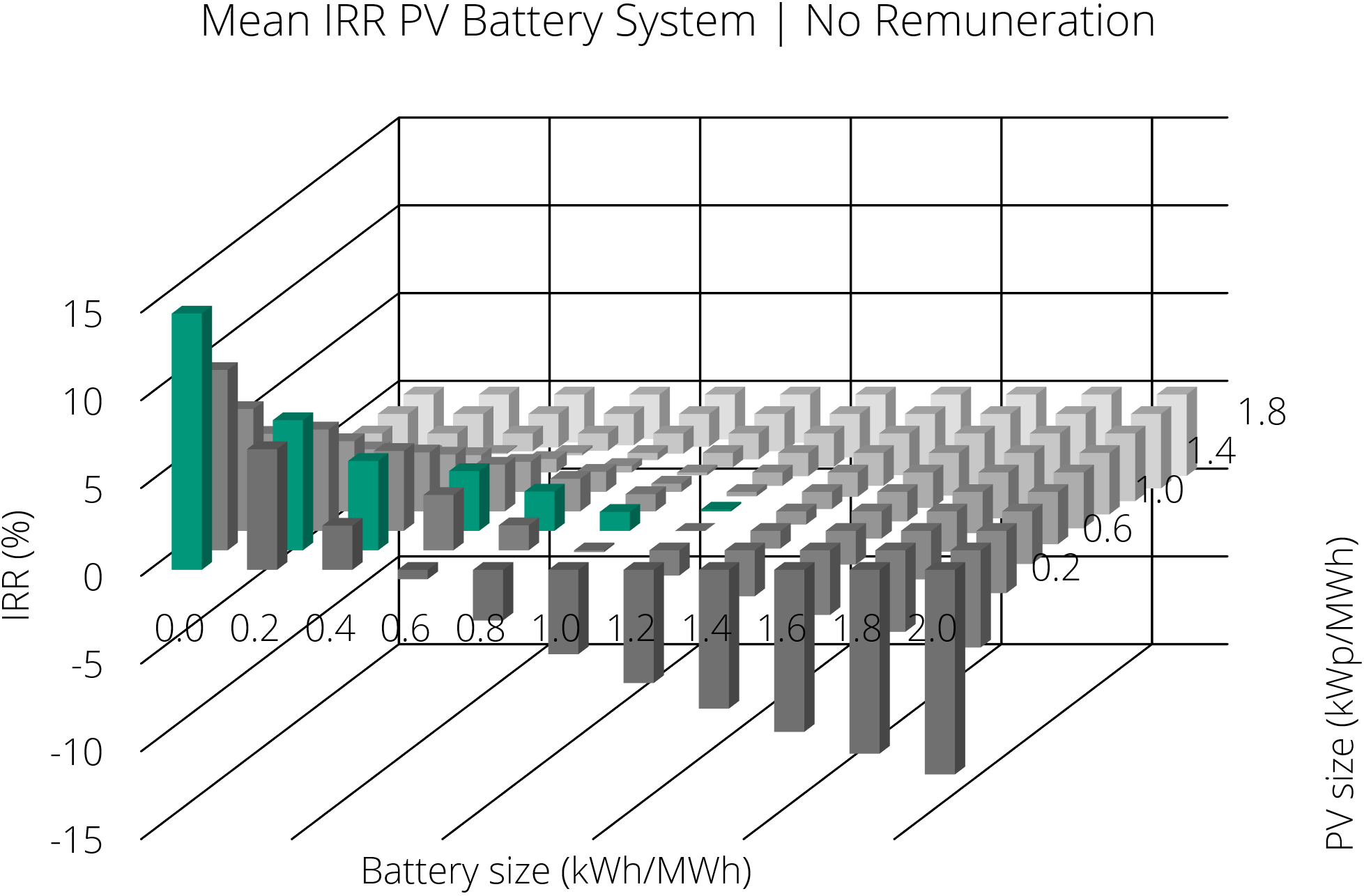}
		\caption{No remuneration for PV feed-in}
		\label{fig:IRR_PVBatt_noR}
	\end{subfigure} 
	\caption{Average IRR of PV battery system, assuming no FIT}
\end{figure}

The IRR was also calculated for the case that no FIT was granted, and excess PV generation electricity can only be sold at an average market remuneration of 40\,EUR/MWh (cp. Fig.~\ref{fig:IRR_PVBatt_Market}, in which only positive maximum IRR values per battery size are marked in green). In this case, average IRR values are lower than with FIT for all system configurations, and even negative for battery sizes beyond 1.4\,kWh/MWh. 

This tendency becomes even stronger if no remuneration for surplus electricity is earned (cp. Fig.~\ref{fig:IRR_PVBatt_noR}). All IRR values are lower than in the market remuneration case, and they are only positive for systems up to 1.2\,kW$_\text{p}$/MWh of PV and up to 1.2\,kWh/MWh of battery capacity. Interestingly, small PV systems without a battery remain profitable, as their benefit is mainly based on saved procurement costs, which is independent of surplus remuneration. 

\begin{figure}[h]
	\centering
	\includegraphics[width=\linewidth]{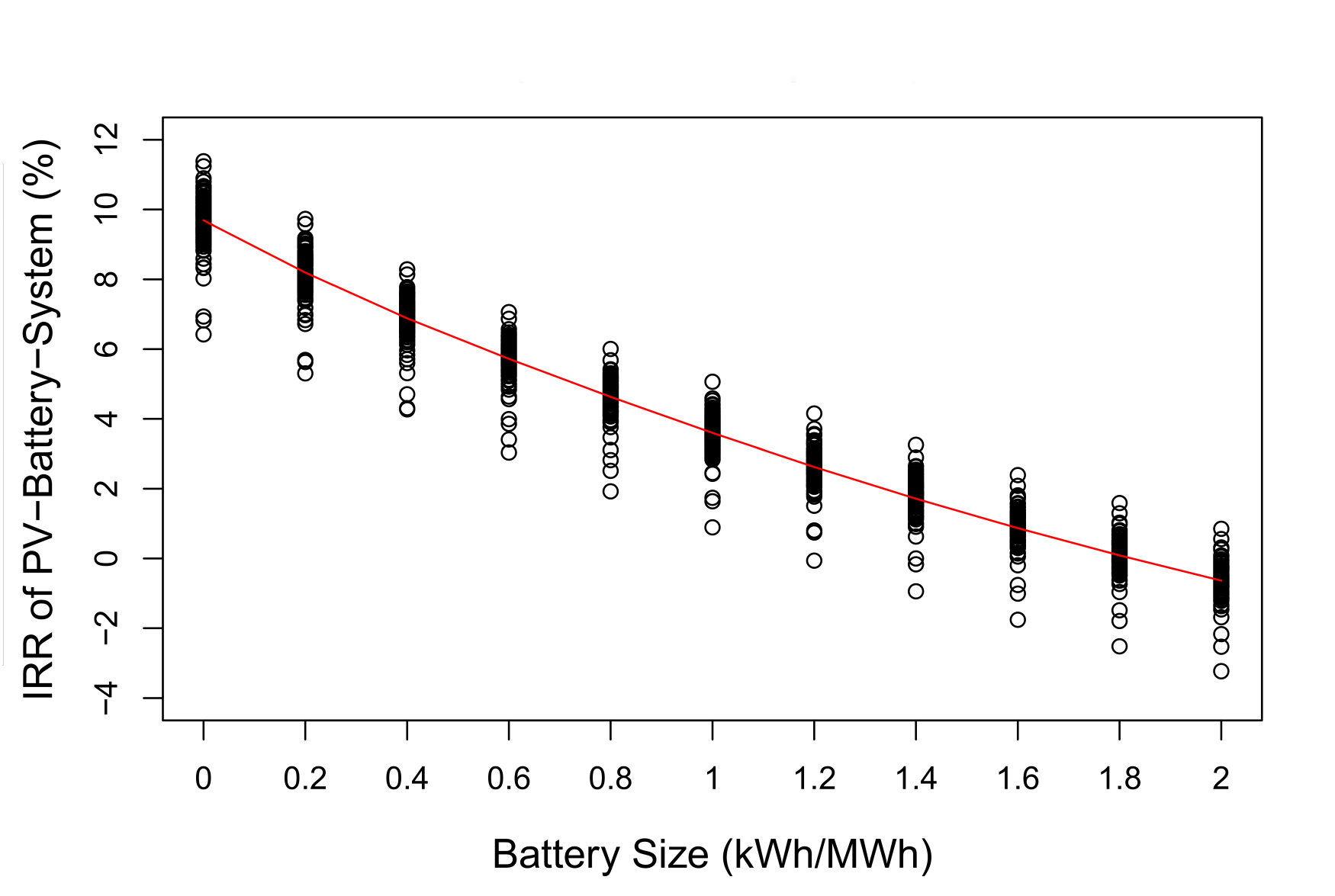}
	\caption{Internal rate of return with FIT as a function of the battery size, for PV plant of 1\,kW$_\text{p}$/MWh plant; solid line: average IRR; dashed lines: average IRR plus/minus one standard deviation}
	\label{fig:ProfitSize}
\end{figure}

The scattering in IRR between properties is rather low, as can be seen in Fig.~\ref{fig:ProfitSize} for the case with FIT. It shows the average and standard deviation of the IRR for systems with a PV size of 1\,kW$_\text{p}$/MWh and different battery sizes. It ranges from 0.3 to 0.7\,\% for most system configurations. Only for systems with a small PV plant and a small battery (resulting in high IRR), the standard deviation is higher (up to 0.9\,\%). The IRR of the system is highly correlated with the respective self-consumption rate of the property, with a Pearson correlation coefficient of $\rho=0.93$ on average (min: 0.79, max: 0.98) for the case with FIT. Correspondingly, most of the variance was explained by summertime consumption (56.5\,\% on average) and type of building (20.3\,\%). The same GLM was used as for SCR, showing that the IRR of nursing homes and museums is on average 0.19 and 0.25\,\% higher than that of schools, while the IRR of conference halls is 0.18\,\% lower. Furthermore, the IRR increases on average by 0.05\,\% for each percent of summertime consumption.

\begin{figure}[h]
	\centering
	\includegraphics[width=\linewidth]{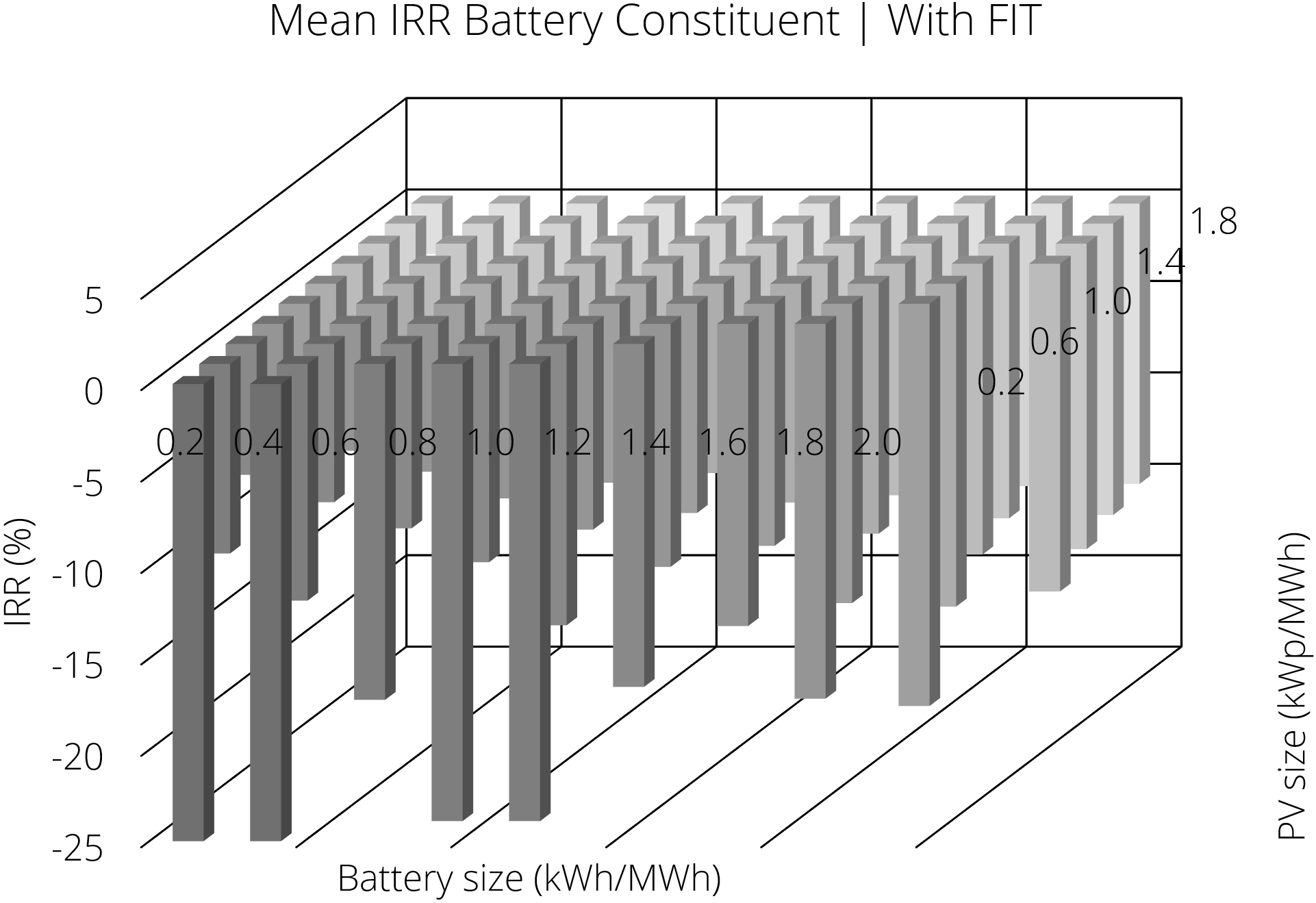}
	\caption{Average IRR of battery investment within PV battery system, assuming FIT}
	\label{fig:IRR_Batt_FIT}
\end{figure}

If the investment into the battery as part of the PV battery system is evaluated separately (setting additional revenues in relation to the battery investment cost), IRR values are considerably lower than for the PV battery system. If current FIT is granted, IRR of the battery is negative for any system sizing, as shown in Fig.~\ref{fig:IRR_Batt_FIT}. In this graph, the ordinate was cut at $-25$\,\%, but some IRR values were much lower than this. The IRR values for the battery constituent are only displayed for system sizing configurations for which the IRR of the complete PV battery system are positive.  

\begin{figure}[h]
		\centering
	\begin{subfigure}[b]{\linewidth}
	\includegraphics[width=\textwidth]{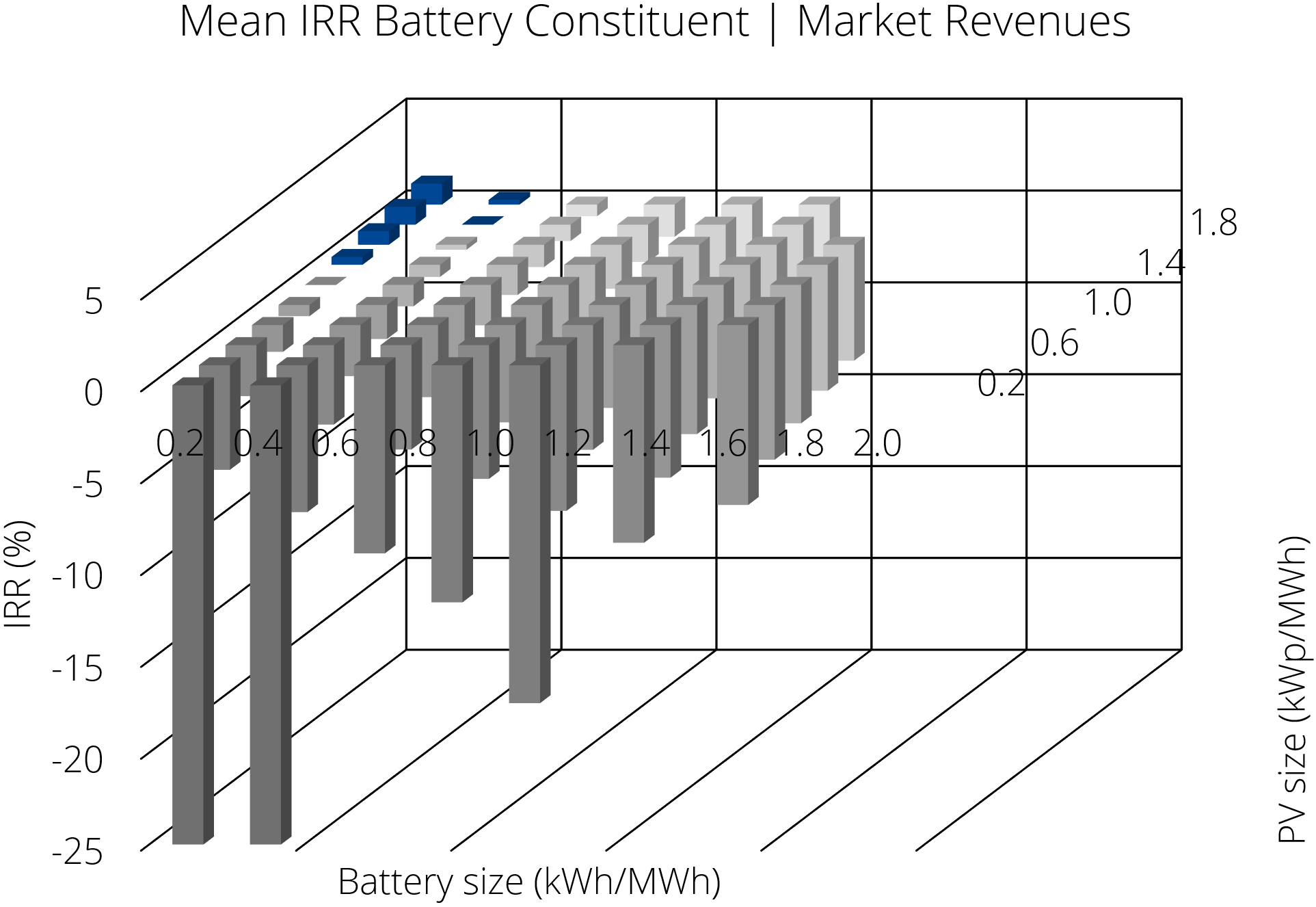}
		\caption{Only market remuneration}
		\label{fig:IRR_Batt_Market}
	\end{subfigure} 

	\begin{subfigure}[b]{\linewidth}
	\includegraphics[width=\textwidth]{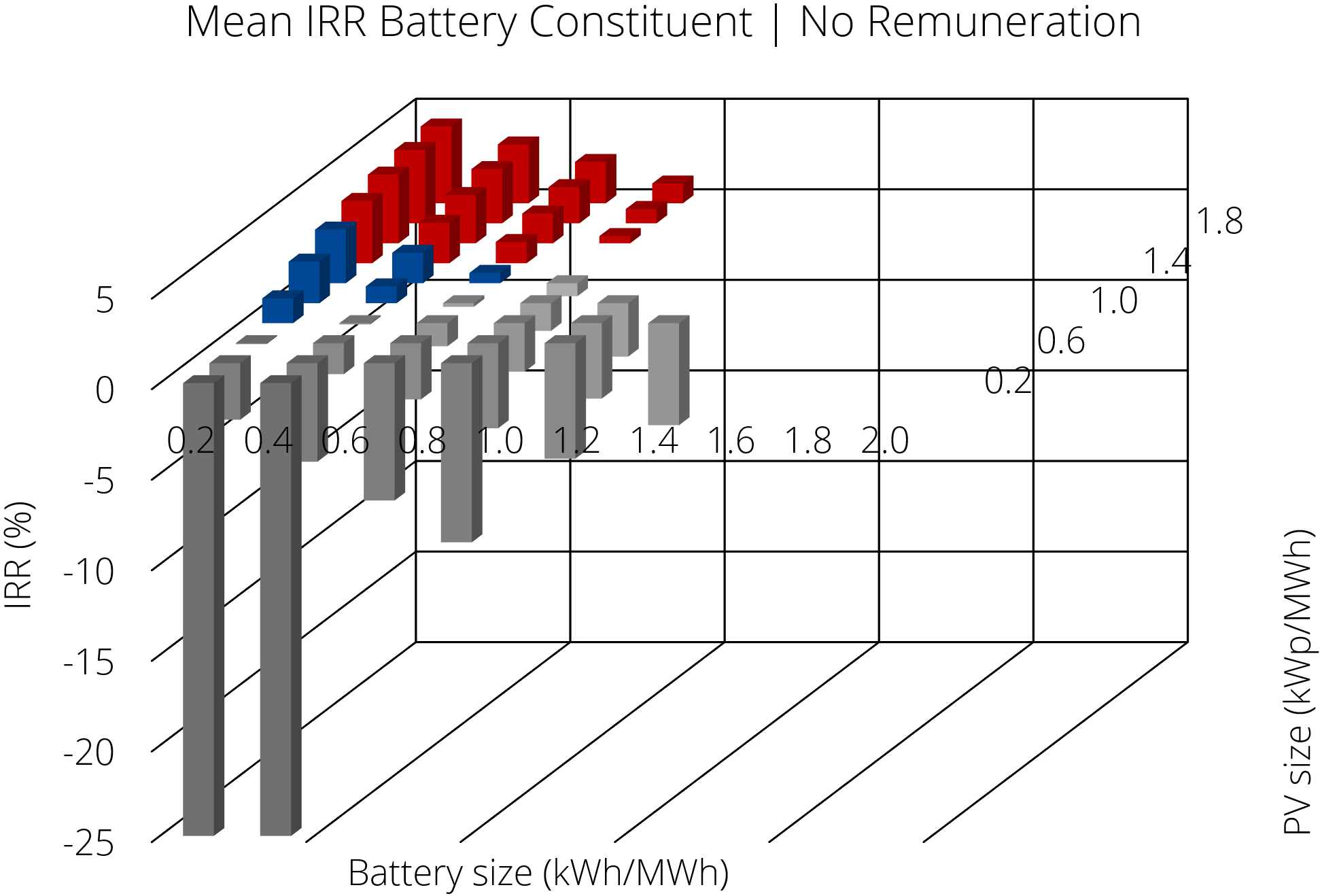}
		\caption{No remuneration for PV feed-in}
		\label{fig:IRR_Batt_noR}
	\end{subfigure} 
	\caption{Average IRR of battery investment within PV battery system, assuming no FIT}
	\label{fig:IRR_Batt_noFIT}
\end{figure}

If only market remuneration is received for the PV surplus, or if no remuneration is earned, batteries' IRR become positive for some system size options, as can be seen in Fig.~\ref{fig:IRR_Batt_noFIT}. In both graphs, blue bars indicate positive values. Values are again displayed only for system configurations with positive IRR of the respective entire system, with the exception of the red bars in Fig.~\ref{fig:IRR_Batt_noR} (no remuneration). These indicate configurations in which the battery provides value to the overall system, with positive IRR values of up to 4.2\,\%, but the overall system IRR is nevertheless negative. 

For the market remuneration case, the battery IRR is only positive for PV systems larger than 1.2\,kW$_\text{p}$/MWh, and for small batteries (0.2 to 0.4\,kWh/MWh). Yet, the highest IRR with 1.1\,\% on average, realized with the largest considered PV plant and the smallest battery considered size, is still quite low, and may not be profitable in the view of many investors. If no surplus electricity is sold, the IRR of the battery is positive for systems with a PV plant of more than 0.6\,kW$_\text{p}$/MWh and up to a battery size of 0.8\,kWh/MWh. The highest IRR values reachable for positive IRR PV battery systems is 3\,\%. 

\begin{figure}[h]
	\centering
	\includegraphics[width=\linewidth]{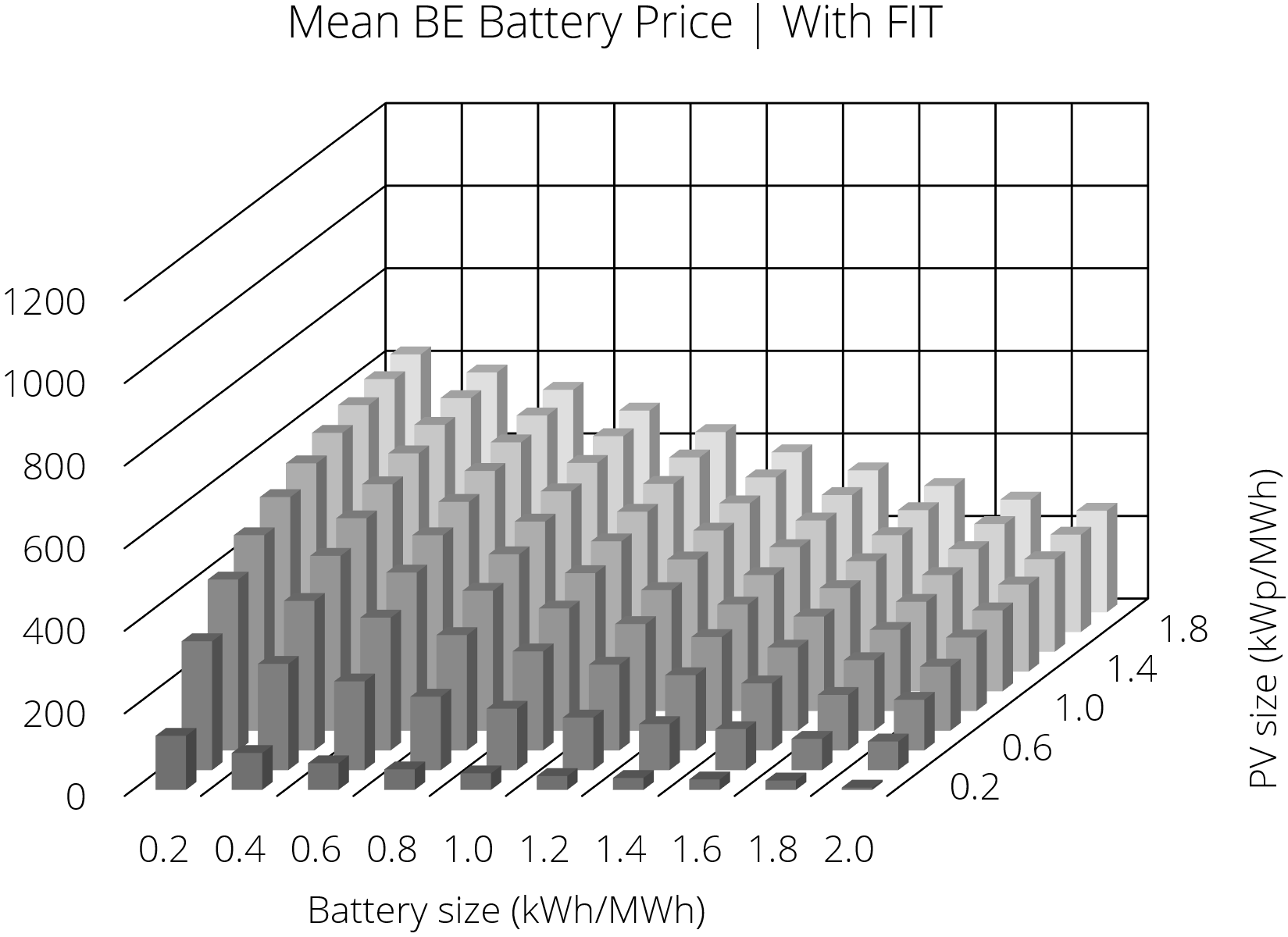}
	\caption{Average break-even price of the battery constituent, assuming FIT}
	\label{fig:BreakEven_FIT}
	\end{figure}

\begin{figure}[h]
	\centering
	\begin{subfigure}[b]{\linewidth}
	\includegraphics[width=\textwidth]{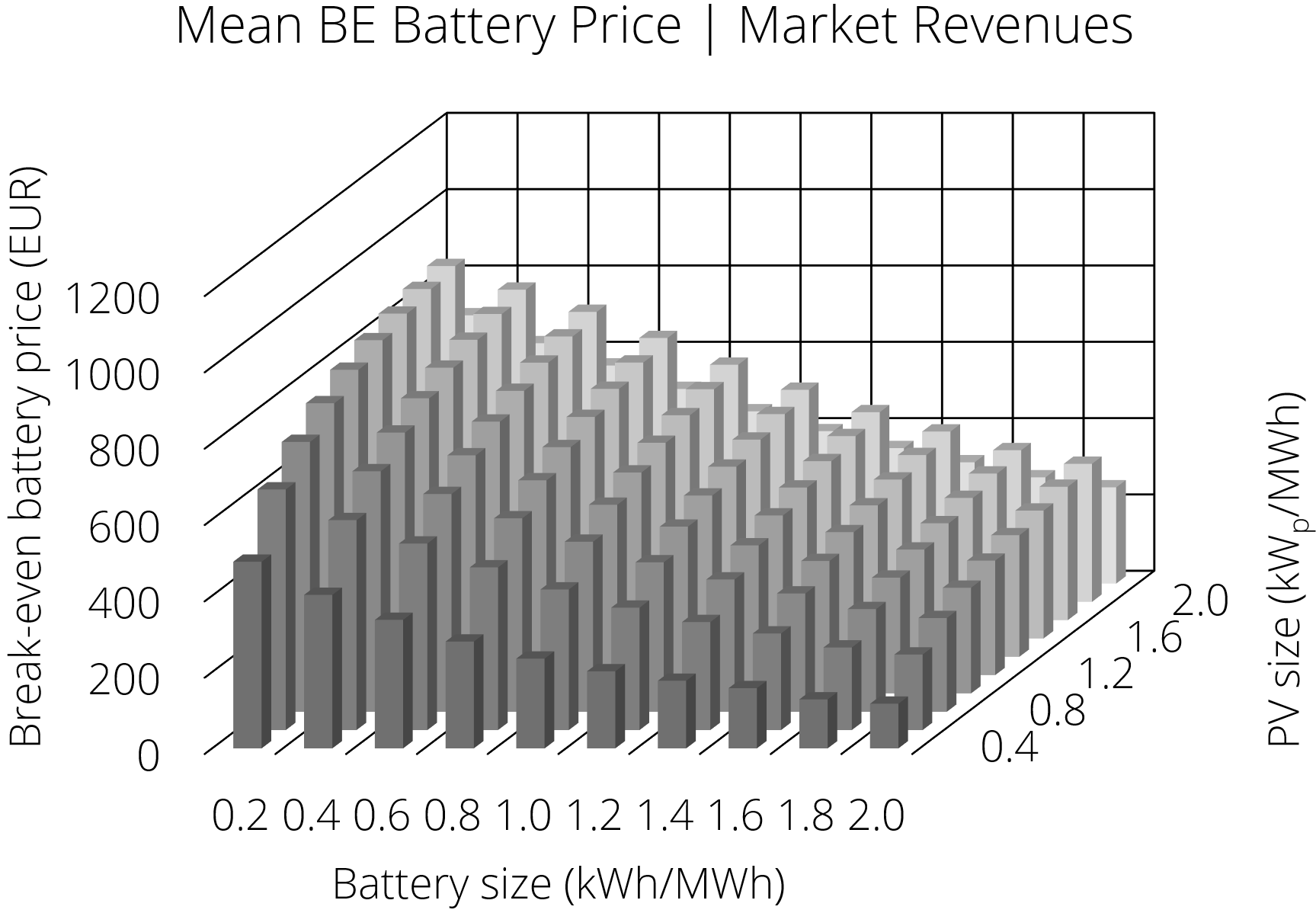}
	\caption{Only market remuneration}
	\label{fig:BreakEven_Market}
	\end{subfigure} 
\begin{subfigure}[b]{\linewidth}
	\includegraphics[width=\textwidth]{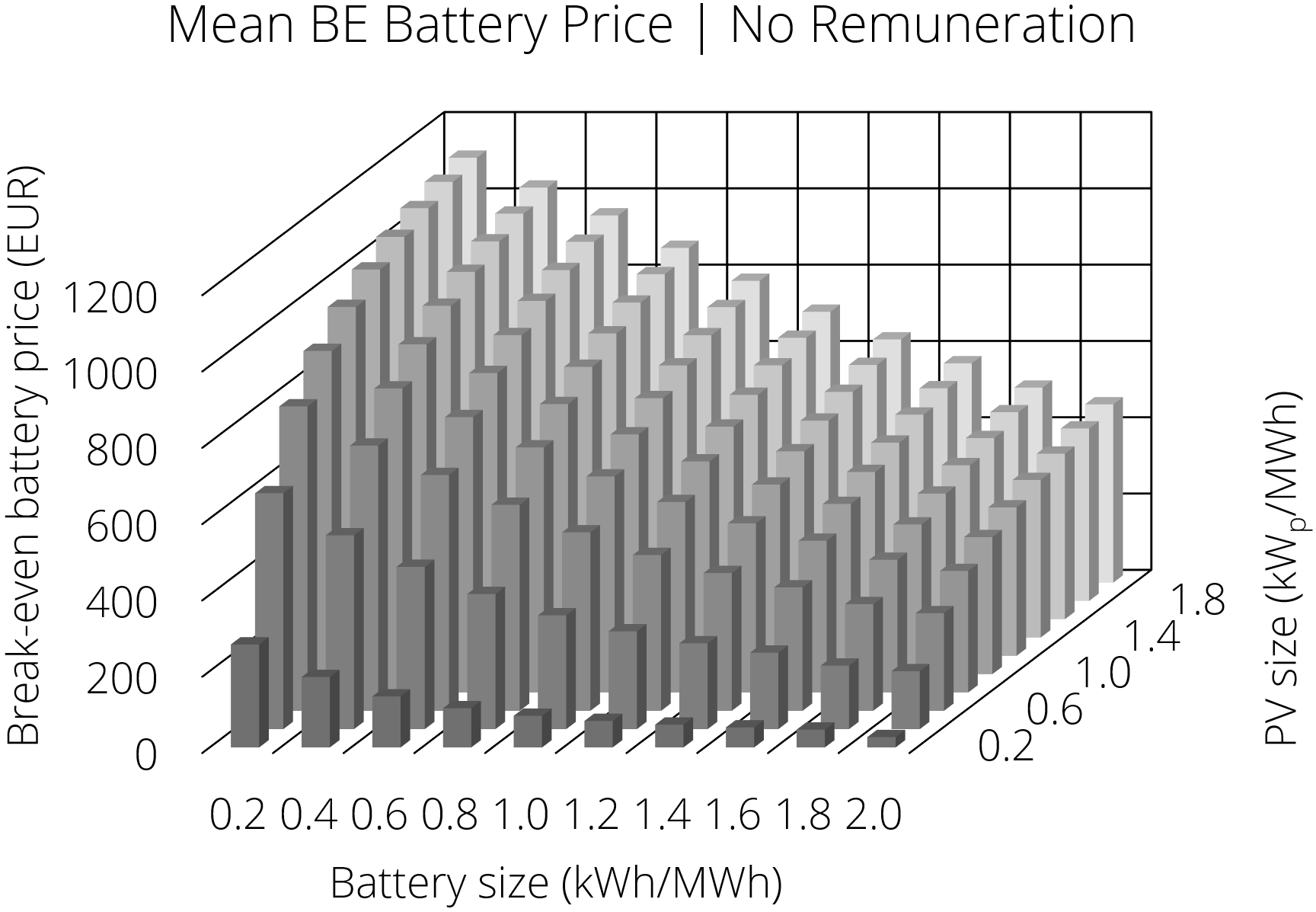}
	\caption{No remuneration for PV feed-in}
	\label{fig:BreakEven_noR}
	\end{subfigure} 
	\caption{Average break-even price of the battery constituent, assuming no FIT}
	\label{fig:BreakEven_noFIT}
\end{figure}

Another interesting indicator if IRR is negative is the battery price (in EUR/kWh of battery capacity) for which a net present value of zero would be achieved. This is here referred to as the break-even (BE) price of the battery. Fig.~\ref{fig:BreakEven_FIT} shows the average break-even prices for all system sizes with FIT, and Fig.~\ref{fig:BreakEven_noFIT} depicts the same for two cases without FIT. BE prices decrease rapidly with increasing battery capacity in all cases, and are lower for small PV plants. The highest break-even price of 731\,EUR/kWh is found for a sports hall with a PV system of 2\,kW$_\text{p}$/MWh and a battery of 0.2\,kWh/MWh. If surplus electricity is sold on the market at a price of 40\,EUR/MWh, an average break-even price of 880\,EUR/kWh is reached for PV systems of 2\,kW$_\text{p}$/MWh and batteries of 0.2\,kWh/MWh. If no surplus electricity remuneration is earned, this value increases to 1\,114\,EUR/kWh.

\begin{figure}[h]
	\centering
	\includegraphics[width=\linewidth]{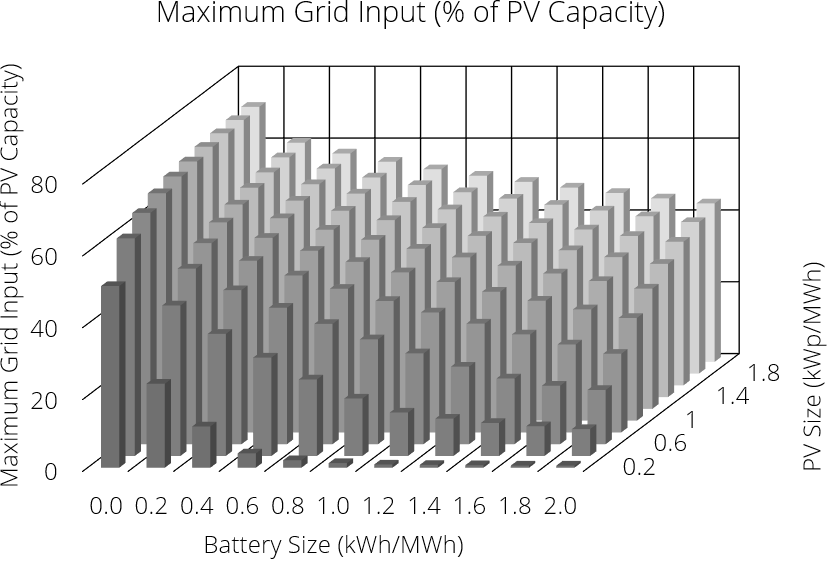}
	\caption{PV feed-in into the grid}
	\label{fig:GridRelief}
\end{figure}

Figure~\ref{fig:GridRelief} shows the maximum grid input for all system sizes (average of all data sets) in percent of installed PV capacity for the model applying $z_2$, i.\,e. the objective function that integrates feed-in minimization. Even with large PV systems, the maximum grid injection is not larger than around 70\,\% of its installed capacity. If the PV plant is combined with a battery, maximum grid input decreases by more than 10\,\% for the first 0.2\,kWh/MWh, and more moderately for additional battery capacity. If cost minimization only ($z_1$) is applied, there is almost no decrease in maximum power feed-in into the grid with increasing battery capacity.

\section{Discussion}\label{sec:discussion}

All results presented in this study are based on data in 15\,min time resolution. It must be noted that every time aggregation reduces the information on actual power peaks of shorter duration to some extent. It has been shown by previous studies that this leads to a systematic overestimation of self-consumption, and an underestimation of losses due to feed-in limits. The authors of \cite{QUOILIN2016} find that the self-sufficiency rate of a 1\,kW$_\text{p}$/MWh PV plant is increased by 1.5\,\% if hourly data is used, compared to data in 1\,min resolution. The study shows that the difference disappears almost completely if the plant is combined with a battery. In \cite{Appen2012}, the losses resulting from a feed-in limitation of 70\,\% of installed PV capacity were modelled for systems without self-consumption. The authors find differences between 1.6 and 2.1\,\% for data in 1\,sec time resolution compared to 15\,min time resolution. Thus, the losses due to the 70\,\% feed-in cap assumed here would probably be lower than in reality. In addition, the grid-friendly model applied (cp. $z_2$, Eq.~\ref{eq:costpower}) represents a BMS with perfect foresight of PV generation and load data. \cite{Bergner2014} find that losses increase from 1 to 3\,\% if PV forecasts are based on the generation of the previous day, and load forecasts on the same weekday of the previous week, compared to perfect foresight. For more realistic results, studies with forecast algorithms would have to be used instead of perfect foresight models, in order ot account for the inherent uncertainty of forecasts.

For reasons of simplicity and to compare the results between properties, it is assumed that there is sufficient roof area for all PV sizes available on each property. In reality, roof shape and area might limit the PV plant size. In addition, the specific prices (EUR/kWh or EUR/kW$_\text{p}$) for both PV and battery are likely to change for different system sizes due to economies of scale, while they are assumed to be constant in this study. The scaling effect would result in higher prices for small PV plants, meaning that the IRR of small systems could be overestimated, here. With increasing battery size, price reduction becomes smaller. 91\,\% of all simulated systems are larger than 20\,kW$_\text{p}$. 

A significant limitation of the economic assessments in this study is that ageing, in particular battery ageing, is  neglected. Sensitivity analyses showed that break-even prices for batteries decrease by 20 to 30\,\% if aging is considered. One weakness of the statistical analysis is the unequal amount of data sets for the different types of buildings. Schools are clearly overrepresented in the data set, while museums, for example, are only represented by two data sets. Model diagnostic showed that there is no pattern in the residuals, when plotted against predicted values, which means that the assumption of linear regression for all variables is appropriate. However, some residuals differ from a normal distribution, which means that some extreme values are not represented properly by the GLM. Including interactions between the predictor variables in the model did not lead to better results.

The higher temporal coincidence of PV generation and electricity demand in municipal properties results in high shares of self-consumption, and high profitability of PV plants alone. Especially small to medium sized PV systems show high internal rates of return. Combining these plants with a battery results only in a moderate increase of self-consumption, but in reduced profitability in all cases assuming feed-in remuneration at the current level in Germany, and in many cases assuming only market revenues or no remuneration at all. Using battery storage, therefore, provides less benefit than it does in residential settings. Besides, the difference between electricity prices and the levelized cost of generating electricity with an own PV plant is smaller for larger consumers, such as municipal buildings, in comparison to private households. Reduced battery prices cannot always improve the IRR for the battery, because the savings per kWh stored are too small. 

Revenue losses due to feed-in power limitations of 70\,\% are found to be well below 1\,\% for all system sizes. For a stricter feed-in power limitation of, for example, 50\,\%, losses account for 1.27\,\% for medium-sized PV systems (1\,kW$_\text{p}$/MWh), and 2.49\,\% for large PV systems (2\,kW$_\text{p}$/MWh) without battery. In the case of the grid-friendly discharging strategy using objective function $z_2$ (Eq.~\ref{eq:costpower}), feed-in losses decrease rapidly with increasing battery capacity. In the case of self-consumption maximization using objective function $z_1$ (cp. Eq.~\ref{eq:cost}), additional battery capacity only results in a small reduction of feed-in losses. 

\section{Conclusions}\label{sec:conclusions}

This study investigated self-consumption rates, self-sufficiency rates and profitability measures for PV battery systems using the load data of 101 municipal properties in 15\,min time resolution. Profitability for all system size configurations were evaluated through the measure of internal rate of return. In addition, the battery prices necessary to bring a PV battery system to a net present value of zero have been calculated. Two different algorithms were used, one that simulates battery management that solely minimizes total electricity cost, and a second one that also keeps power injection to the grid as low as possible.

It was found that (i) IRR and break-even prices for batteries greatly depend on the respective system size;  (ii) the highest IRR is reached with small to medium sized PV plants without battery storage, and (iii) for smaller PV plants, the combination with a battery storage is not profitable.  However, (iv) small batteries can be profitable if they are combined with a large PV plant in the case that no feed-in tariff is granted, which applies, e.\,g., to PV plants at the end of their support period.

 On the level of the self-consumption and self-sufficiency rates, it was found that (v) SCR and SSR of PV battery systems installed on municipal properties are significantly higher than for residential buildings in the case of small batteries; (vi) for medium sized batteries (1\,kWh/MWh) the differences are negligible and for large batteries (2\,kWh/MWh) self-consumption of residential buildings is even higher.

  Regarding the grid interaction, it was found that (vii) if the objective function is cost minimization only, PV feed-in power cannot be reduced significantly by the battery, because the battery is already fully charged before the peak in PV generation is reached. However, (viii) if grid injection is added to the objective function to be minimized, even small batteries lead to considerable reductions of feed-in power.

 Finally, the findings suggest that (ix) differences in self-consumption and profitability are low between the different types of municipal buildings, and the share of summertime consumption has the greatest impact on both self-consumption and profitability, while (x) the share of daytime consumption impacts profitability only for systems with large PV plants and small batteries.

In summary, the high profitability of PV systems on municipal properties ensures an economic operation in most cases for the PV plant sizes investigated here. One benefit of storage, while not being economically advantageous in the large majorities of cases considered here, is that even small batteries can help shaving feed-in power peaks significantly if an optimized charging and discharging strategy is used. 

\bibliographystyle{elsarticle-num} 
\bibliography{References}







\appendix

\newpage
    \begingroup
\let\clearpage\relax 
\onecolumn 
	\section{Numerical values of Figs. 1, 5, 6, 7 and  9--13 (additional information, not included in the accepted manuscript)}
\endgroup

\begin{table*}[h]
	\centering
	\begin{tabular}{|r|r|r|r|r|r|r|r|r|r|r|r|}
		\hline
 \multicolumn{12}{|c|}{Mean Self-Cconsumption Rate (in \%)} \\
		\hline
  \multicolumn{12}{|l|}{PV Size (kW$_\text{p}$/MWh)}  \\
		\hline 
\multicolumn{2}{|l|}{} & \textbf{0.2} & \textbf{0.4} & \textbf{0.6} & \textbf{0.8} & \textbf{1.0} & \textbf{1.2} &\textbf{1.4} & \textbf{1.6} & \textbf{1.8} & \textbf{2.0} \\
		\hline
\parbox[t]{2mm}{\multirow{11}{*}{\rotatebox[origin=c]{90}{Battery Size (kWh/MWh)}}} & \textbf{0.0} & 85.6 & 68.2 & 55.5 & 46.6 & 40.2 & 35.4 & 31.6 & 28.6 & 26.1 & 24.0 \\
\cline{2-12}
& \textbf{0.2} & 93.3 & 77.0 & 63.1 & 53.0 & 45.7 & 40.2 & 35.9 & 32.4 & 29.6 & 27.2 \\
\cline{2-12}
& \textbf{0.4} & 96.2 & 82.7 & 68.8 & 58.0 & 50.1 & 44.1 & 39.4 & 35.6 & 32.5 & 29.9 \\
\cline{2-12}
& \textbf{0.6} & 97.1 & 86.4 & 73.2 & 62.1 & 53.7 & 47.4 & 42.4 & 38.4 & 35.0 & 32.2 \\
\cline{2-12}
& \textbf{0.8} & 97.4 & 88.4 & 76.1 & 65.0 & 56.4 & 49.8 & 44.6 & 40.4 & 36.9 & 34.0 \\
\cline{2-12}
& \textbf{1.0} & 97.5 & 89.5 & 77.8 & 66.8 & 58.2 & 51.4 & 46.1 & 41.8 & 38.3 & 35.3 \\
\cline{2-12}
& \textbf{1.2} & 97.6 & 90.2 & 78.8 & 68.0 & 59.3 & 52.5 & 47.2 & 42.8 & 39.2 & 36.1 \\
\cline{2-12}
& \textbf{1.4} & 97.6 & 90.7 & 79.6 & 68.8 & 60.1 & 53.2 & 47.8 & 43.4 & 39.8 & 36.7 \\
\cline{2-12}
& \textbf{1.6} & 97.7 & 91.1 & 80.2 & 69.4 & 60.6 & 53.7 & 48.3 & 43.9 & 40.2 & 37.1 \\
\cline{2-12}
& \textbf{1.8} & 97.7 & 91.4 & 80.6 & 69.8 & 61.0 & 54.1 & 48.6 & 44.2 & 40.5 & 37.4 \\
\cline{2-12}
& \textbf{2.0} & 97.7 & 91.7 & 81.0 & 70.2 & 61.4 & 54.4 & 48.9 & 44.4 & 40.7 & 37.6 \\
		\hline
		\end{tabular}
	\caption{Numerical values of Fig. \ref{fig:SCR}}
\end{table*}

\begin{table*}[h]
	\centering
	\begin{tabular}{|r|r|r|r|r|r|r|r|r|r|r|r|}
		\hline
		\multicolumn{12}{|c|}{Mean Self-Sufficiency Rate (in \%)} \\
		\hline
		\multicolumn{12}{|l|}{PV Size (kW$_\text{p}$/MWh)}  \\
		\hline 
\multicolumn{2}{|l|}{} & \textbf{0.2} & \textbf{0.4} & \textbf{0.6} & \textbf{0.8} & \textbf{1.0} & \textbf{1.2} &\textbf{1.4} & \textbf{1.6} & \textbf{1.8} & \textbf{2.0} \\
		\hline
\parbox[t]{2mm}{\multirow{11}{*}{\rotatebox[origin=c]{90}{Battery Size (kWh/MWh)}}} & \textbf{0.0} & 16.9 & 26.9 & 32.9 & 36.9 & 39.7 & 41.9 & 43.7 & 45.2 & 46.4 & 47.4 \\
\cline{2-12}
& \textbf{0.2} & 18.4 & 30.4 & 37.4 & 41.9 & 45.2 & 47.7 & 49.6 & 51.2 & 52.6 & 53.7 \\
\cline{2-12}
&\textbf{0.4} & 19.0 & 32.7 & 40.8 & 45.9 & 49.5 & 52.3 & 54.5 & 56.3 & 57.8 & 59.1 \\
\cline{2-12}
& \textbf{0.6} & 19.2 & 34.1 & 43.4 & 49.1 & 53.1 & 56.2 & 58.6 & 60.6 & 62.3 & 63.7 \\
\cline{2-12}
& \textbf{0.8} & 19.2 & 34.9 & 45.1 & 51.4 & 55.7 & 59.0 & 61.7 & 63.9 & 65.7 & 67.2 \\
\cline{2-12}
& \textbf{1.0} & 19.3 & 35.4 & 46.1 & 52.8 & 57.5 & 61.0 & 63.8 & 66.1 & 68.1 & 69.7 \\
\cline{2-12}
& \textbf{1.2} & 19.3 & 35.6 & 46.7 & 53.8 & 58.6 & 62.3 & 65.2 & 67.7 & 69.7 & 71.4 \\
\cline{2-12}
& \textbf{1.4} & 19.3 & 35.8 & 47.2 & 54.4 & 59.4 & 63.1 & 66.1 & 68.7 & 70.8 & 72.6 \\
\cline{2-12}
& \textbf{1.6} & 19.3 & 36.0 & 47.5 & 54.8 & 59.9 & 63.7 & 66.8 & 69.4 & 71.5 & 73.3 \\
\cline{2-12}
& \textbf{1.8} & 19.3 & 36.1 & 47.8 & 55.2 & 60.3 & 64.1 & 67.2 & 69.8 & 72.0 & 73.9 \\
\cline{2-12}
& \textbf{2.0} & 19.3 & 36.2 & 48.0 & 55.5 & 60.6 & 64.5 & 67.6 & 70.2 & 72.4 & 74.3 \\
		\hline
\end{tabular}
\caption{Numerical values of Fig. \ref{fig:SSR}}
\end{table*}

\begin{table*}[h]
	\centering
	\begin{tabular}{|r|r|r|r|r|r|r|r|r|r|r|r|}
		\hline
		\multicolumn{12}{|c|}{Mean IRR PV Battery System | With FIT (in \%)} \\
		\hline
		\multicolumn{12}{|l|}{PV Size (kW$_\text{p}$/MWh)}  \\
		\hline 
		\multicolumn{2}{|l|}{} & \textbf{0.2} & \textbf{0.4} & \textbf{0.6} & \textbf{0.8} & \textbf{1.0} & \textbf{1.2} &\textbf{1.4} & \textbf{1.6} & \textbf{1.8} & \textbf{2.0} \\
		\hline
		\parbox[t]{2mm}{\multirow{11}{*}{\rotatebox[origin=c]{90}{Battery Size (kWh/MWh)}}} & \textbf{0.0} & 16.3 & 14.0 & 12.2 & 10.8 & 9.7 & 8.8 & 8.1 & 7.6 & 7.1 & 6.7 \\
		\cline{2-12}
		& \textbf{0.2} & 7.3 & 9.4 & 9.3 & 8.8 & 8.2 & 7.7 & 7.2 & 6.8 & 6.4 & 6.1 \\
		\cline{2-12}
		& \textbf{0.4} & 2.7 & 6.3 & 7.1 & 7.1 & 6.9 & 6.6 & 6.3 & 6.0 & 5.8 & 5.5  \\
		\cline{2-12}
		& \textbf{0.6} & -0.5 & 3.9 & 5.3 & 5.7 & 5.7 & 5.6 & 5.5 & 5.3 & 5.2 & 5.0 \\
		\cline{2-12}
		& \textbf{0.8} & -2.9 & 2.0 & 3.7 & 4.4 & 4.6 & 4.7 & 4.7 & 4.6 & 4.5 & 4.4 \\
		\cline{2-12}
		& \textbf{1.0} & -4.8 & 0.3 & 2.3 & 3.2 & 3.6 & 3.8 & 3.9 & 3.9 & 3.9 & 3.8  \\
		\cline{2-12}
		& \textbf{1.2} & -6.4 & -1.1 & 1.1 & 2.1 & 2.6 & 2.9 & 3.1 & 3.2 & 3.2 & 3.2  \\
		\cline{2-12}
		& \textbf{1.4} & -7.9 & -2.3 & 0.0 & 1.1 & 1.7 & 2.1 & 2.3 & 2.5 & 2.6 & 2.6  \\ 
		\cline{2-12}
		& \textbf{1.6} & -9.2 & -3.4 & -1.0 & 0.2 & 0.9 & 1.3 & 1.6 & 1.8 & 1.9 & 2.0  \\
		\cline{2-12}
		& \textbf{1.8} & -10.4 & -4.4 & -1.9 & -0.7 & 0.1 & 0.6 & 0.9 & 1.2 & 1.3 & 1.5  \\
		\cline{2-12}
		& \textbf{2.0} & -11.6 & -5.3 & -2.7 & -1.4 & -0.6 & -0.1 & 0.3 & 0.5 & 0.8 & 0.9  \\
		\hline
	\end{tabular}
	\caption{Numerical values of Fig. \ref{fig:IRR_PVBatt_FIT}}
\end{table*}

\begin{table*}[h]
	\centering
	\begin{tabular}{|r|r|r|r|r|r|r|r|r|r|r|r|}
		\hline
		\multicolumn{12}{|c|}{Mean IRR PV battery system | Market Revenues (in \%)} \\
		\hline
		\multicolumn{12}{|l|}{PV Size (kW$_\text{p}$/MWh)}  \\
		\hline 
		\multicolumn{2}{|l|}{} & \textbf{0.2} & \textbf{0.4} & \textbf{0.6} & \textbf{0.8} & \textbf{1.0} & \textbf{1.2} &\textbf{1.4} & \textbf{1.6} & \textbf{1.8} & \textbf{2.0} \\
		\hline
		\parbox[t]{2mm}{\multirow{11}{*}{\rotatebox[origin=c]{90}{Battery Size (kWh/MWh)}}} & \textbf{0.0} & 15.3 & 11.9 & 9.2 & 7.3 & 5.9 & 4.7 & 3.8 & 3.0 & 2.3 & 1.8 \\
		\cline{2-12}
		& \textbf{0.2} & 7.0 & 8.2 & 7.3 & 6.2 & 5.1 & 4.3 & 3.5 & 2.8 & 2.2 & 1.7 \\
		\cline{2-12}
		& \textbf{0.4} & 2.6 & 5.6 & 5.7 & 5.1 & 4.4 & 3.7 & 3.1 & 2.5 & 2.0 & 1.6  \\
		\cline{2-12}
		& \textbf{0.6} & -0.5 & 3.5 & 4.2 & 4.0 & 3.6 & 3.1 & 2.6 & 2.2 & 1.7 & 1.3  \\
		\cline{2-12}
		& \textbf{0.8} & -2.9 & 1.7 & 2.9 & 3.0 & 2.8 & 2.4 & 2.1 & 1.7 & 1.4 & 1.0  \\
		\cline{2-12}
		& \textbf{1.0} & -4.8 & 0.1 & 1.6 & 2.0 & 1.9 & 1.7 & 1.5 & 1.2 & 0.9 & 0.6  \\
		\cline{2-12}
		& \textbf{1.2} & -6.4 & -1.3 & 0.5 & 1.0 & 1.1 & 1.0 & 0.8 & 0.6 & 0.4 & 0.2  \\
		\cline{2-12}
		& \textbf{1.4} & -7.9 & -2.5 & -0.6 & 0.1 & 0.3 & 0.3 & 0.2 & 0.0 & -0.1 & -0.3  \\
		\cline{2-12}
		& \textbf{1.6} & -9.2 & -3.6 & -1.5 & -0.8 & -0.5 & -0.4 & -0.5 & -0.5 & -0.7 & -0.8  \\
		\cline{2-12}
		& \textbf{1.8} & -10.5 & -4.5 & -2.4 & -1.6 & -1.2 & -1.1 & -1.1 & -1.1 & -1.2 & -1.3 \\
		\cline{2-12}
		& \textbf{2.0} & -11.6 & -5.4 & -3.2 & -2.3 & -1.9 & -1.7 & -1.6 & -1.6 & -1.7 & -1.8  \\
\hline
	\end{tabular}
	\caption{Numerical values of Fig. \ref{fig:IRR_PVBatt_Market}}
\end{table*}

\begin{table*}[h]
	\centering
	\begin{tabular}{|r|r|r|r|r|r|r|r|r|r|r|r|}
		\hline
		\multicolumn{12}{|c|}{Mean IRR PV Battery System | No Remuneration (in \%)} \\
		\hline
		\multicolumn{12}{|l|}{PV Size (kW$_\text{p}$/MWh)}  \\
		\hline 
		\multicolumn{2}{|l|}{} & \textbf{0.2} & \textbf{0.4} & \textbf{0.6} & \textbf{0.8} & \textbf{1.0} & \textbf{1.2} &\textbf{1.4} & \textbf{1.6} & \textbf{1.8} & \textbf{2.0} \\
		\hline
		\parbox[t]{2mm}{\multirow{11}{*}{\rotatebox[origin=c]{90}{Battery Size (kWh/MWh)}}} & \textbf{0.0} & 14.6 & 10.3 & 6.9 & 4.4 & 2.4 & 0.8 & -0.6 & -1.8 & -2.8 & -3.8   \\ 
		\cline{2-12}
		& \textbf{0.2} & 6.9 & 7.4 & 5.8 & 4.0 & 2.4 & 1.1 & -0.2 & -1.2 & -2.2 & -3.1  \\
		\cline{2-12}
		& \textbf{0.4} & 2.5 & 5.1 & 4.6 & 3.4 & 2.1 & 1.0 & -0.1 & -1.0 & -1.9 & -2.7   \\ 
		\cline{2-12}
		& \textbf{0.6} & -0.5 & 3.2 & 3.4 & 2.7 & 1.7 & 0.8 & -0.1 & -1.0 & -1.8 & -2.5  \\
		\cline{2-12}
		& \textbf{0.8} & -2.9 & 1.4 & 2.2 & 1.9 & 1.1 & 0.4 & -0.4 & -1.2 & -1.9 & -2.6   \\ 
		\cline{2-12}
		& \textbf{1.0} & -4.8 & -0.1 & 1.1 & 1.0 & 0.5 & -0.2 & -0.8 & -1.5 & -2.1 & -2.7  \\
		\cline{2-12}
		& \textbf{1.2} & -6.5 & -1.4 & 0.0 & 0.1 & -0.3 & -0.8 & -1.3 & -1.9 & -2.5 & -3.0  \\ 
		\cline{2-12}
		& \textbf{1.4} & -7.9 & -2.6 & -1.0 & -0.7 & -1.0 & -1.4 & -1.9 & -2.4 & -2.9 & -3.4  \\
		\cline{2-12}
		& \textbf{1.6} & -9.2 & -3.7 & -1.9 & -1.6 & -1.7 & -2.0 & -2.4 & -2.9 & -3.3 & -3.8  \\ 
		\cline{2-12}
		& \textbf{1.8} & -10.5 & -4.6 & -2.8 & -2.3 & -2.3 & -2.6 & -3.0 & -3.4 & -3.8 & -4.2  \\ 
		\cline{2-12}
		& \textbf{2.0} & -11.7 & -5.5 & -3.5 & -3.0 & -3.0 & -3.2 & -3.5 & -3.9 & -4.2 & -4.6  \\
		\hline
	\end{tabular}
	\caption{Numerical values of Fig. \ref{fig:IRR_PVBatt_noR}}
\end{table*}

\begin{table*}[h]
	\centering
	\begin{tabular}{|r|r|r|r|r|r|r|r|r|r|r|r|}
		\hline
		\multicolumn{12}{|c|}{Mean IRR Battery Constituent | With FIT (in \%)} \\
		\hline
		\multicolumn{12}{|l|}{PV Size (kW$_\text{p}$/MWh)}  \\
		\hline 
		\multicolumn{2}{|l|}{} & \textbf{0.2} & \textbf{0.4} & \textbf{0.6} & \textbf{0.8} & \textbf{1.0} & \textbf{1.2} &\textbf{1.4} & \textbf{1.6} & \textbf{1.8} & \textbf{2.0} \\
		\hline
		\parbox[t]{2mm}{\multirow{11}{*}{\rotatebox[origin=c]{90}{Battery Size (kWh/MWh)}}} & \textbf{0.2} &-98.5 & -10.4 & -7.2 & -5.7 & -4.8 & -4.1 & -3.6 & -3.2 & -3.0 & -2.8 \\ 
		\cline{2-12}
		& \textbf{0.4} & -153.2 & -12.9 & -8.7 & -6.9 & -5.9 & -5.1 & -4.6 & -4.1 & -3.8 & -3.6  \\
		\cline{2-12}
		& \textbf{0.6} &  & -18.4 & -10.1 & -8.1 & -6.9 & -6.1 & -5.5 & -5.0 & -4.7 & -4.4  \\
		\cline{2-12}
		&  \textbf{0.8} &  & -30.0 & -11.9 & -9.5 & -8.3 & -7.4 & -6.7 & -6.2 & -5.8 & -5.5  \\
		\cline{2-12}
		&  \textbf{1.0} &  & -66.1 & -15.4 & -11.3 & -9.8 & -8.8 & -8.1 & -7.5 & -7.1 & -6.7  \\
		\cline{2-12}
		&  \textbf{1.2} &  &  & -18.8 & -13.3 & -11.4 & -10.4 & -9.6 & -8.9 & -8.5 & -8.1 \\
		\cline{2-12}
		&  \textbf{1.4} &  &  &  & -16.5 & -13.2 & -12.0 & -11.1 & -10.4 & -9.9 & -9.5 \\
		\cline{2-12}
		&  \textbf{1.6} &  &  &  & -20.5 & -16.4 & -13.7 & -12.7 & -11.9 & -11.3 & -10.9  \\
		\cline{2-12}
		& \textbf{1.8} &  &  &  &  & -22.0 & -17.7 & -15.9 & -15.0 & -14.4 & -13.8  \\
		\cline{2-12}
		&  \textbf{2.0} &  &  &  &  &  &  & -17.9 & -16.7 & -15.9 & -15.3  \\
		\hline
	\end{tabular}
	\caption{Numerical values of Fig. \ref{fig:IRR_Batt_FIT}}
\end{table*}

\begin{table*}[h]
	\centering
	\begin{tabular}{|r|r|r|r|r|r|r|r|r|r|r|r|}
		\hline
		\multicolumn{12}{|c|}{Mean IRR Battery Constituent | Market Revenues (in \%)} \\
		\hline
		\multicolumn{12}{|l|}{PV Size (kW$_\text{p}$/MWh)}  \\
		\hline 
		\multicolumn{2}{|l|}{} & \textbf{0.2} & \textbf{0.4} & \textbf{0.6} & \textbf{0.8} & \textbf{1.0} & \textbf{1.2} &\textbf{1.4} & \textbf{1.6} & \textbf{1.8} & \textbf{2.0} \\
		\hline
		\parbox[t]{2mm}{\multirow{11}{*}{\rotatebox[origin=c]{90}{Battery Size (kWh/MWh)}}} & \textbf{0.2} &-50.2 & -5.7 & -2.8 & -1.5 & -0.6 & 0.0 & 0.4 & 0.7 & 1.0 & 1.1\\
		\cline{2-12}
		&  \textbf{0.4} & -104.8 & -8.0 & -4.3 & -2.8 & -1.9 & -1.2 & -0.7 & -0.3 & 0.0 & 0.3 \\
		\cline{2-12}
		&  \textbf{0.6} &  & -10.2 & -5.7 & -3.9 & -2.9 & -2.2 & -1.6 & -1.2 & -0.9 & -0.6  \\
		\cline{2-12}
		&  \textbf{0.8} &  & -12.9 & -7.3 & -5.3 & -4.2 & -3.4 & -2.9 & -2.4 & -2.0 & -1.8  \\
		\cline{2-12}
		&  \textbf{1.0} &  & -18.4 & -9.0 & -6.8 & -5.6 & -4.8 & -4.2 & -3.7 & -3.3 & -3.0  \\
		\cline{2-12}
		&  \textbf{1.2} &  &  & -10.8 & -8.3 & -7.0 & -6.2 & -5.5 & -5.0 & -4.6 & -4.3 \\
		\cline{2-12}
		&  \textbf{1.4} &  &  &  & -9.8 & -8.4 & -7.5 & -6.9 & -6.3 &  &   \\
		\cline{2-12}
		&  \textbf{1.6} &  &  &  &  &  &  &  &  &  & \\
		\cline{2-12}
		&  \textbf{1.8} &  &  &  &  &  &  &  &  &  & \\
		\cline{2-12}
		&  \textbf{2.0} &  &  &  &  &  &  &  &  &  & \\
		\hline
	\end{tabular}
	\caption{Numerical values of Fig. \ref{fig:IRR_Batt_Market}}
\end{table*}

\begin{table*}[h]
	\centering
	\begin{tabular}{|r|r|r|r|r|r|r|r|r|r|r|r|}
		\hline
		\multicolumn{12}{|c|}{Mean IRR Battery Constituent | No Remuneration (in \%)} \\
		\hline
		\multicolumn{12}{|l|}{PV Size (kW$_\text{p}$/MWh)}  \\
		\hline 
		\multicolumn{2}{|l|}{} & \textbf{0.2} & \textbf{0.4} & \textbf{0.6} & \textbf{0.8} & \textbf{1.0} & \textbf{1.2} &\textbf{1.4} & \textbf{1.6} & \textbf{1.8} & \textbf{2.0} \\
		\hline
		\parbox[t]{2mm}{\multirow{11}{*}{\rotatebox[origin=c]{90}{Battery Size (kWh/MWh)}}} & \textbf{0.2} &-34.1 & -3.1 & -0.1 & 1.4 & 2.3 & 3.0 & 3.4 & 3.8 & 4.0 & 4.2 \\
			\cline{2-12}
		& \textbf{0.4} & -75.9 & -5.4 & -1.7 & -0.1 & 0.9 & 1.7 & 2.2 & 2.7 & 3.0 & 3.3 \\		
			\cline{2-12}
		& \textbf{0.6} &  & -7.6 & -3.1 & -1.3 & -0.2 & 0.6 & 1.2 & 1.6 & 2.0 & 2.3 \\
			\cline{2-12}
		& \textbf{0.8} &  & -9.9 & -4.7 & -2.7 & -1.5 & -0.7 &  & 0.4 & 0.8 & 1.1 \\
			\cline{2-12}
		& \textbf{1.0} &  &  & -6.4 & -4.2 & -2.9 &  &  &  &  &  \\
			\cline{2-12}
		& \textbf{1.2} &  &  &  & -5.6 &  &  &  &  &  &  \\
			\cline{2-12}
		& \textbf{1.4} &  &  &  &  &  &  &  &  &  &  \\
			\cline{2-12}
		& \textbf{1.6} &  &  &  &  &  &  &  &  &  &  \\
			\cline{2-12}
		& \textbf{1.8} &  &  &  &  &  &  &  &  &  &  \\
			\cline{2-12}
		& \textbf{2.0} &  &  &  &  &  &  &  &  &  & \\
		\hline
	\end{tabular}
	\caption{Numerical values of Fig. \ref{fig:IRR_Batt_noR}}
\end{table*}

\begin{table*}[h]
	\centering
	\begin{tabular}{|r|r|r|r|r|r|r|r|r|r|r|r|}
		\hline
		\multicolumn{12}{|c|}{Mean BE Battery Price | With FIT (in EUR)} \\
		\hline
		\multicolumn{12}{|l|}{PV Size (kW$_\text{p}$/MWh)}  \\
		\hline 
		\multicolumn{2}{|l|}{} & \textbf{0.2} & \textbf{0.4} & \textbf{0.6} & \textbf{0.8} & \textbf{1.0} & \textbf{1.2} &\textbf{1.4} & \textbf{1.6} & \textbf{1.8} & \textbf{2.0} \\
		\hline
		\parbox[t]{2mm}{\multirow{11}{*}{\rotatebox[origin=c]{90}{Battery Size (kWh/MWh)}}} & \textbf{0.2} &131.0 & 312.3 & 413.9 & 473.6 & 518.1 & 552.3 & 578.2 & 597.6 & 612.7 & 624.9   \\ 
			\cline{2-12}
		& \textbf{0.4} & 89.4 & 257.4 & 362.2 & 423.6 & 466.7 & 501.4 & 528.4 & 549.8 & 566.8 & 581.1   \\
		\cline{2-12}
		& \textbf{0.6} & 64.6 & 215.1 & 321.7 & 383.2 & 425.7 & 458.8 & 485.9 & 507.4 & 524.6 & 539.0   \\
		\cline{2-12}
		& \textbf{0.8} & 49.6 & 178.0 & 279.4 & 339.0 & 379.6 & 411.1 & 436.6 & 457.5 & 474.6 & 488.6   \\
			\cline{2-12}
		& \textbf{1.0} & 40.1 & 148.9 & 240.0 & 296.1 & 334.0 & 363.2 & 387.1 & 406.8 & 423.1 & 436.7   \\
		\cline{2-12}
		& \textbf{1.2} & 33.7 & 127.5 & 207.7 & 258.3 & 292.9 & 319.4 & 341.4 & 359.7 & 374.9 & 387.7   \\ 
			\cline{2-12}
		& \textbf{1.4} & 29.0 & 111.6 & 182.3 & 226.9 & 257.9 & 281.5 & 301.3 & 318.2 & 332.1 & 344.0   \\ 
			\cline{2-12}
		& \textbf{1.6} & 25.4 & 99.5 & 162.4 & 201.5 & 228.7 & 249.9 & 267.3 & 282.7 & 295.4 & 306.3  \\
			\cline{2-12}
		& \textbf{1.8} & 22.7 & 76.0 & 134.6 & 170.9 & 196.6 & 216.8 & 233.7 & 248.7 & 261.7 & 273.0  \\
			\cline{2-12}
		& \textbf{2.0} & 5.5 & 69.5 & 122.8 & 155.3 & 178.0 & 195.9 & 210.8 & 224.3 & 236.0 & 246.3  \\
		\hline
	\end{tabular}
	\caption{Numerical values of Fig. \ref{fig:BreakEven_FIT}}
\end{table*}

\begin{table*}[h]
	\centering
	\begin{tabular}{|r|r|r|r|r|r|r|r|r|r|r|r|}
		\hline
		\multicolumn{12}{|c|}{Mean BE Battery Price | Market Revenues (in EUR)} \\
		\hline
		\multicolumn{12}{|l|}{PV Size (kW$_\text{p}$/MWh)}  \\
		\hline 
		\multicolumn{2}{|l|}{} & \textbf{0.2} & \textbf{0.4} & \textbf{0.6} & \textbf{0.8} & \textbf{1.0} & \textbf{1.2} &\textbf{1.4} & \textbf{1.6} & \textbf{1.8} & \textbf{2.0} \\
		\hline
		\parbox[t]{2mm}{\multirow{11}{*}{\rotatebox[origin=c]{90}{Battery Size (kWh/MWh)}}} & \textbf{0.2} & 214.0 & 489.2 & 630.6 & 707.1 & 760.5 & 800.2 & 829.7 & 851.3 & 867.8 & 880.1   \\ 
			\cline{2-12}
		& \textbf{0.4} & 146.2 & 402.3 & 550.1 & 630.4 & 683.6 & 725.4 & 757.5 & 782.7 & 802.3 & 818.2   \\
			\cline{2-12}
		& \textbf{0.6} & 105.9 & 336.7 & 488.9 & 570.7 & 623.9 & 664.3 & 697.1 & 723.0 & 743.5 & 759.9   \\
			\cline{2-12}
		& \textbf{0.8} & 81.3 & 279.9 & 426.3 & 506.9 & 558.5 & 597.5 & 628.7 & 654.3 & 675.1 & 691.4   \\
			\cline{2-12}
		& \textbf{1.0} & 65.7 & 235.3 & 368.5 & 445.6 & 494.5 & 531.1 & 560.8 & 585.1 & 605.2 & 621.5   \\
			\cline{2-12}
		& \textbf{1.2} & 55.1 & 202.1 & 320.9 & 391.6 & 437.1 & 470.7 & 498.3 & 521.2 & 540.1 & 555.7  \\
			\cline{2-12}
		& \textbf{1.4} & 47.4 & 177.2 & 283.2 & 346.6 & 388.1 & 418.5 & 443.5 & 464.9 & 482.4 & 497.0  \\
		\cline{2-12}
		& \textbf{1.6} & 41.6 & 157.9 & 253.2 & 310.0 & 347.2 & 374.7 & 397.0 & 416.6 & 432.8 & 446.4  \\ 
			\cline{2-12}
		& \textbf{1.8} & 37.0 & 128.4 & 216.5 & 269.0 & 303.4 & 329.1 & 350.0 & 368.5 & 384.2 & 397.4   \\
			\cline{2-12}
		& \textbf{2.0} & 18.4 & 117.1 & 197.9 & 245.7 & 276.9 & 300.0 & 318.7 & 335.3 & 349.6 & 361.8  \\
		\hline
	\end{tabular}
	\caption{Numerical values of Fig. \ref{fig:BreakEven_Market}}
\end{table*}

\begin{table*}[h]
	\centering
	\begin{tabular}{|r|r|r|r|r|r|r|r|r|r|r|r|}
		\hline
		\multicolumn{12}{|c|}{Mean BE Battery Price | No Remuneration (in EUR)} \\
		\hline
		\multicolumn{12}{|l|}{PV Size (kW$_\text{p}$/MWh)}  \\
		\hline 
		\multicolumn{2}{|l|}{} & \textbf{0.2} & \textbf{0.4} & \textbf{0.6} & \textbf{0.8} & \textbf{1.0} & \textbf{1.2} &\textbf{1.4} & \textbf{1.6} & \textbf{1.8} & \textbf{2.0} \\
		\hline
		\parbox[t]{2mm}{\multirow{11}{*}{\rotatebox[origin=c]{90}{Battery Size (kWh/MWh)}}} & \textbf{0.2} & 270.0 & 618.4 & 798.1 & 895.2 & 962.7 & 1012.9 & 1050.2 & 1077.6 & 1098.4 & 1113.9  \\ 
			\cline{2-12}
		& \textbf{0.4} & 184.5 & 508.0 & 695.2 & 797.0 & 864.6 & 917.6 & 958.5 & 990.5 & 1015.4 & 1035.5 \\ 
			\cline{2-12}
		& \textbf{0.6} & 133.7 & 425.4 & 618.1 & 721.9 & 789.4 & 840.7 & 882.4 & 915.4 & 941.5 & 962.5  \\
			\cline{2-12}
		& \textbf{0.8} & 102.7 & 354.2 & 540.0 & 642.4 & 708.1 & 757.7 & 797.4 & 830.0 & 856.6 & 877.5  \\
					\cline{2-12}
		& \textbf{1.0} & 83.0 & 298.3 & 467.9 & 566.3 & 628.7 & 675.5 & 713.5 & 744.5 & 770.2 & 791.1 \\
					\cline{2-12}
		& \textbf{1.2} & 69.5 & 256.5 & 408.4 & 499.3 & 557.8 & 600.8 & 636.3 & 665.5 & 689.8 & 709.9  \\
					\cline{2-12}
		& \textbf{1.4} & 59.8 & 224.9 & 361.2 & 443.4 & 497.0 & 536.2 & 568.5 & 596.0 & 618.6 & 637.5  \\
					\cline{2-12}
		& \textbf{1.6} & 52.5 & 200.4 & 323.5 & 397.7 & 446.3 & 482.0 & 511.0 & 536.4 & 557.4 & 575.0  \\
					\cline{2-12}
		& \textbf{1.8} & 46.7 & 166.5 & 279.8 & 348.0 & 392.6 & 425.5 & 451.8 & 475.0 & 494.7 & 511.1 \\
					\cline{2-12}
		& \textbf{2.0} & 27.1 & 151.7 & 255.9 & 318.6 & 359.4 & 389.3 & 413.1 & 434.1 & 452.1 & 467.3  \\
		\hline
	\end{tabular}
	\caption{Numerical values of Fig. \ref{fig:BreakEven_noR}}
\end{table*}

\begin{table*}[h]
	\centering
	\begin{tabular}{|r|r|r|r|r|r|r|r|r|r|r|r|}
		\hline
		\multicolumn{12}{|c|}{Maximum Grid Input (\% of PV Capacity)} \\
		\hline
		\multicolumn{12}{|l|}{PV Size (kW$_\text{p}$/MWh)}  \\
		\hline 
		\multicolumn{2}{|l|}{} & \textbf{0.2} & \textbf{0.4} & \textbf{0.6} & \textbf{0.8} & \textbf{1.0} & \textbf{1.2} &\textbf{1.4} & \textbf{1.6} & \textbf{1.8} & \textbf{2.0} \\
		\hline
		\parbox[t]{2mm}{\multirow{11}{*}{\rotatebox[origin=c]{90}{Battery Size (kWh/MWh)}}} & \textbf{0.0} & 50.6 & 60.6 & 64.4 & 66.6 & 68.0 & 68.9 & 69.7 & 70.2 & 70.6 & 71.0 \\
					\cline{2-12}
		& \textbf{0.2} & 23.4 & 41.9 & 48.9 & 52.7 & 55.2 & 56.9 & 58.3 & 59.3 & 60.2 & 61.0 \\
					\cline{2-12}
		& \textbf{0.4} & 11.5 &  34.0 & 42.9 & 47.8 & 50.9 & 53.1 & 54.7 & 56.0 & 57.1 & 58.0 \\
					\cline{2-12}
		& \textbf{0.6} & 4.0 & 27.4 & 38.0 & 43.7 & 47.3 & 49.9 & 51.9 & 53.4 & 54.6 & 55.7 \\
					\cline{2-12}
		& \textbf{0.8} & 2.1 & 21.3 & 33.5 & 40.0 & 44.2 & 47.1 & 49.3 & 51.1 & 52.5 & 53.6 \\
					\cline{2-12}
		& \textbf{1.0} & 1.3 & 16.0 & 29.2 & 36.6 & 41.3 & 44.6 & 47.1 & 49.0 & 50.5 & 51.8 \\
					\cline{2-12}
		& \textbf{1.2} & 0.9 & 12.1 & 25.3 & 33.4 & 38.6 & 42.2 & 44.9 & 47.0 & 48.7 & 50.1 \\
					\cline{2-12}
		& \textbf{1.4} & 0.8 & 10.4 & 21.6 & 30.3 & 35.9 & 39.9 & 42.9 & 45.2 & 47.0 & 48.5 \\
					\cline{2-12}
		& \textbf{1.6} & 0.7 & 9.2 & 18.3 & 27.3 & 33.4 & 37.7 & 40.9 & 43.4 & 45.4 & 47.0 \\
					\cline{2-12}
		& \textbf{1.8} & 0.6 & 8.3 & 16.3 & 24.5 & 31.0 & 35.6 & 39.0 & 41.6 & 43.8 & 45.5 \\
					\cline{2-12}
		& \textbf{2.0} & 0.6 & 7.5 & 15.1 & 21.9 & 28.6 & 33.5 & 37.1 & 40.0 & 42.2 & 44.1 \\
		\hline
	\end{tabular}
	\caption{Numerical values of Fig. \ref{fig:GridRelief}}
\end{table*}

\end{document}